\pdfoutput=1
\documentclass[cernpreprint, atlasdraft=false, texlive=2016, UKenglish, texmf]{atlasdoc}
 
\usepackage{atlaspackage}
\usepackage{float}
\floatstyle{plaintop}
\restylefloat{table}
 
\usepackage{atlasbiblatex}
 
\usepackage{atlasphysics}
\usepackage{multirow}
\usepackage{amsmath}
\newcolumntype{H}{>{\setbox0=\hbox\bgroup}c<{\egroup}@{}}
\newcolumntype{Z}{>{\setbox0=\hbox\bgroup}c<{\egroup}@{\hspace*{-\tabcolsep}}}

\graphicspath{{logos/}{figures/}}
 
\usepackage{ANA-STDM-2019-18-PAPER-defs}
 
\AtlasTitle{Determination of the parton distribution functions of the proton from ATLAS measurements of differential $W^\pm$ and $Z$ boson production in association with jets}
 
\AtlasVersion{3.0}
\AtlasAbstract{
 
This article presents a new set of proton parton distribution
functions, ATLASepWZVjet20, produced in an analysis at
next-to-next-to-leading order in QCD.
The new data sets considered are
the measurements of $W^+$ and  $W^-$ boson and $Z$ boson production in
association with jets in $pp$ collisions at $\sqrt{s} = 8~\mathrm{TeV}$ performed by the ATLAS experiment at the LHC
with integrated luminosities of $20.2~\mathrm{fb}^{-1}$ and $19.9~\mathrm{fb}^{-1}$,
respectively. The analysis also considers the ATLAS measurements of
differential $W^{\pm}$ and $Z$ boson production at $\sqrt{s} = 7~\mathrm{TeV}$ with an
integrated luminosity of $4.6~\mathrm{fb}^{-1}$ and deep-inelastic-scattering data
from $e^{\pm}p$ collisions at the HERA accelerator.  An improved
determination of the sea-quark densities at high Bjorken $x$ is shown,
while confirming a strange-quark density similar in size to the up- and
down-sea-quark densities in the range $x \lesssim 0.02$ found by previous
ATLAS analyses.
}
 
\author{The ATLAS Collaboration}
 
\AtlasRefCode{STDM-2019-18}
\AtlasDate{\today}

\AtlasJournal{JHEP}
\PreprintIdNumber{CERN-EP-2020-237}
\AtlasJournalRef{JHEP 07 (2021) 223}
\AtlasDOI{10.1007/JHEP07(2021)223}
\hypersetup{pdftitle={ATLAS document},pdfauthor={The ATLAS Collaboration}}
 
\begin{document}
 
\maketitle

% The next lines are included from the .//sections/Introduction.tex input file
\section{Introduction}
 
Precise knowledge of the content of colliding protons, the parton distribution functions (PDFs), is a necessary ingredient for accurate predictions of both Standard Model (SM) and Beyond Standard Model (BSM) cross sections at the Large Hadron Collider (LHC).
In order to determine the PDFs to the required
precision, data covering a wide range of negative squared four-momentum transfer (denoted by $Q^2$) and Bjorken $x$, the fraction of the proton's longitudinal momentum carried by the parton initiating the interaction, is required.
This is facilitated by combining data from multiple experiments and measurements of various processes to constrain the $x$-dependence and flavour decomposition of the PDFs.
While deep inelastic scattering (DIS) data from lepton--hadron collisions typically deliver the best constraints by utilising the lepton as a direct probe of the substructure of the hadron, a hadron--hadron experiment can provide valuable additional insight by introducing new processes which further distinguish contributions from different partons and span kinematic regions at higher $Q^2$.
 
Precision measurements by the HERA collaborations \cite{herapdf20} of neutral current (NC) and charged current (CC) cross sections in $e^\pm p$ scattering constrain PDFs such that
the HERA DIS data alone provide sufficient information to determine the PDF set referred to as HERAPDF2.0. However, they do have limitations. For example, they cannot distinguish quark flavour between the down-type sea quarks, $\bar{d}$ and $\bar{s}$.
Global PDF analyses \cite{ABMP16,CT18,MMHT14,NNPDF31} use a range of data from other experiments together with the HERA data for further constraining power. For example, additional information about quarks and antiquarks at mid- to high-$x$ comes from fixed-target DIS experiments, as well as measurements of $W$ and $Z$ boson production from the Tevatron and LHC experiments.
 
More information about high-$x$ quarks would be advantageous since a large fraction of the fixed-target DIS data is in a kinematic region where non-perturbative effects, such as those from higher twist, are important and must be computed from phenomenological models \cite{Botje:1999dj,Alekhin:2003qq}. In many PDF analyses, tight cuts are applied to these data to avoid those effects. Furthermore, the interpretation of DIS data using deuteron or heavier nuclei as targets is subject to uncertain nuclear corrections. The $W^\pm$ asymmetry measurements performed using \ppbar{} collisions at Tevatron are free from these uncertainties, but there have historically been tensions between the results of the CDF \cite{0901.2169} and D\O\ \cite{WasymD0elchannel} collaborations, discussed in further detail by the MSTW group in Ref.~\cite{0901.0002} and the CTEQ group in Ref.~\cite{Gao:2013xoa}.
 
Precision measurements from the ATLAS detector at the LHC, together with data from the HERA experiments, have been interpreted previously in a next-to-next-to-leading-order (NNLO) QCD analysis, resulting in the ATLASepWZ16 PDF set \cite{STDM-2012-20}.
Differential $W$ and $Z$ boson\footnote{Where $Z$ boson production is written, this refers to $Z/\gamma^*$ boson production.} cross-section measurements at $\rts = 7$~\TeV{} were used, thereby allowing the strange content of the sea to be fitted, rather than assumed to be a
fixed fraction of the light sea as is required when fitting HERA inclusive data alone.
It was found that these additional data were significantly better described by a
strange sea unsuppressed relative to the up- and down-quark sea at $x \lesssim 0.05$, in contradiction to previous assumptions based on dimuon production data from muon-neutrino CC DIS with associated charm-quark production \cite{E866}.
The finding of an unsuppressed strange PDF in this kinematic region is supported by the ATLAS measurement of $W$ boson production in association with a charm quark ($W + c$) at 7~\TeV{} \cite{STDM-2012-14}; however, a recent analysis of CMS $W+c$ data at 7 and 13~\TeV{} \cite{CMS-SMP-17-014} has found a suppressed strange-quark density relative to the light sea, which is potentially in tension with these ATLAS findings.
 
Data on the production of a vector boson in association with jets at the LHC provides a novel source of input to PDF determination that is sensitive to partons at higher $x$ and $Q^2$ than can be accessed by $W$ and $Z$ boson data alone, thereby yielding a data set complementary to the inclusive $W,Z$ boson measurements \cite{2013arXiv1304.2424A}.
The tree-level production modes of a vector boson in association with jets ($V$~+~jets) have either quark--antiquark initial states with gluon radiation, or quark--gluon initial states.
The process is therefore already sensitive to the gluon density of the proton at leading order in quantum chromodynamics (QCD), while providing constraints on the quark distributions in a similar way to inclusive production of a vector boson.
 
This paper presents a PDF analysis including data on $W^{\pm}$~+~jets and $Z$~+~jets
production collected in \pp{} collisions at
$\sqrt{s} = 8~\TeV$ by the ATLAS Collaboration \cite{STDM-2016-14,STDM-2016-11} in combination
with the previous inclusive $W$ and $Z$ measurements at $\sqrt{s} = 7~\TeV$ \cite{STDM-2012-20} and the inclusive combined HERA data \cite{herapdf20}.
The PDF fit is performed at NNLO in perturbative QCD, made possible by recent theoretical developments for vector-boson production in association with one jet \cite{WjetNNLO, ZjetNNLO}, and accounts for the correlation of systematic uncertainties between data sets.
The resulting PDF set is called \emph{ATLASepWZVjet20}.
% End of text imported from the .//sections/Introduction.tex input file
 
% The next lines are included from the .//sections/Theory.tex input file
\section{Input data sets}
 
The final combined $e^{\pm}p$
cross-section measurements at HERA~\cite{herapdf20}
cover the kinematic range of  $Q^2$
from $0.045$~\GeV$^2$ to $50\,000$~\GeV$^2$ and
of Bjorken $x$ from $0.65$ down to $6\times 10^{-7}$.
Data below $x = 10^{-5}$ are excluded from this analysis by requiring $Q^2 > 10~\GeV{}^2$, motivated by the previously observed poorer fit quality in the excluded kinematic region compared to the rest of the HERA data \cite{herapdf20}.
Possible explanations for this include the need for resummation corrections at low $x$ \cite{NLLx} and the impact of higher-twist corrections at low $Q^2$.
For the final HERA data set, there are 169 correlated sources of uncertainty.
Total uncertainties are below 1.5$\%$ over the $Q^2$ range of $10 < Q^2 <500$~\GeV$^2$ and below 3$\%$ up to $Q^2=3000$~\GeV$^2$.
 
The ATLAS $W$ and $Z$ differential cross sections are based on data recorded during \pp{} collisions with $\rts = 7~\TeV$, and a total integrated luminosity of 4.6~\ifb, in the electron and muon boson-decay channels \cite{STDM-2012-20}.
The $W^{\pm}$ differential
cross sections are measured
as functions of the $W$-decay lepton
pseudorapidity, $\eta_\ell$, split into $W^{+}$ and $W^{-}$ cross sections. The experimental precision is between $0.6\%$ and $1.0\%$.
Double-differential distributions of the dilepton rapidity, $y_{\ell\ell}$, in $Z$ boson decays are measured in three mass ranges: $46<m_{\ell\ell}<66$~\GeV, $66<m_{\ell\ell}<116$~\GeV{}
and $116<m_{\ell\ell}<150$~\GeV{} in central ($|y_{\ell\ell}| < 2.4$) and forward ($1.2 < |y_{\ell\ell}| < 3.6$) rapidity selections, with an
experimental precision of up to $0.4\%$ for central rapidity and $2.3\%$ for forward rapidity.
The integrated luminosity of the data set used for the 7~\TeV{} $W$ and $Z$ cross-section measurements is known to within $1.8\%$.
There are a total of 131 sources of correlated systematic uncertainty across the $W$ and $Z$ data sets~\cite{STDM-2012-20}.
These data were used for the ATLASepWZ16 fit in a format in which the measurements of the electron and
muon decay channels were combined, whereas for the PDF sets presented in this article the data before this combination is used.
This choice was made because the uncombined data retain the physical origin of the sources of correlated uncertainties,
thereby allowing these sources to be treated as correlated with those in other data sets.
 
The ATLAS $W^{\pm}$~+~jets differential cross sections are based on data recorded during \pp{} collisions with $\rts = 8~\TeV$
and a total integrated luminosity of 20.2~\ifb,
in the electron decay channel only \cite{STDM-2016-14}. Each event contains at least one jet with transverse momentum $\pt > 30~\GeV{}$ and rapidity $\lvert y\rvert < 4.4$, where jets are defined using the anti-\kt{} algorithm \cite{Cacciari:2008gp,Fastjet} with a radius parameter $R = 0.4$.
The spectrum used is the transverse momentum of the $W$ boson (\WPT), in the range $25<\WPT<800~\GeV{}$, chosen because it provides the most constraining power. This is split into $W^{+}$ and $W^{-}$ cross sections, which have large correlations that are fully considered.
The experimental uncertainty ranges from 8.2\% to 22.1\%~\cite{STDM-2016-14}.
There are 50 sources of correlated systematic uncertainty common to the $W^{+}$ and $W^{-}$ spectra, as well as three sources of uncorrelated uncertainties related to data statistics, background Monte Carlo (MC) simulation statistics and the statistical uncertainty of the data-driven multijet background estimation.
Full information about the statistical bin-to-bin correlations in data, induced by the unfolding process, is available for each $W$~+~jets spectrum.
 
The ATLAS $Z$~+~jets double-differential cross sections are also based on data recorded during \pp{} collisions with $\rts = 8~\TeV$.
The total integrated luminosity of this data is 19.9~\ifb, and the $Z \rightarrow \ee$ decay channel is used \cite{STDM-2016-11}. The measurement is performed as a function of the absolute rapidity of inclusive anti-\kt{} $R=0.4$ jets, \yjets,
for several bins of the transverse momentum within $25~\GeV < \pt^{\mathrm{jet}} < 1050~\GeV$.
The experimental precision ranges from 4.7\% to 37.1\%.
There are 42 sources of correlated systematic uncertainty and two sources of uncorrelated uncertainty related to the data and background MC simulation statistics.
 
The integrated luminosity of the data set used for the $W$~+~jets and $Z$~+~jets cross-section measurements is known to within $1.9\%$.
Systematic uncertainties which contribute significantly, such as from the jet energy scale, are treated as correlated across data sets if they correspond to the same physical source. More details of the correlation model used in this analysis are given in \cref{app:correlations}.
 
\section{Fit framework}
\label{sec:theory}
 
This determination of proton PDFs uses the
xFitter framework, v2.0.1~\cite{HERAFitter,herapdf20,Aaron:2009kv}. This program interfaces to theoretical calculations directly or uses fast interpolation grids to make theoretical predictions for the considered processes. The program MINUIT~\cite{minuit} is used for the minimisation of the PDF fit. The results are cross-checked with an
independent fit framework~\cite{zeusfitter}.
 
For the DIS processes, coefficient functions with massless quarks are calculated at NNLO in QCD as implemented in QCDNUM v17-01-13~\cite{Botje:2010ay}.
The contributions of heavy quarks are calculated in the general-mass variable-flavour-number scheme of Refs.~\cite{Thorne1,Thorne2,Thorne:2006qt}.
The renormalisation and factorisation scales for the DIS processes are taken as
$\muR=\muF=\sqrt{Q^2}$.
 
For the differential $W$ and $Z$ boson cross sections, the theoretical framework is the same as that used in the ATLASepWZ16 analysis of Ref.~\cite{STDM-2012-20}. The xFitter package uses outputs from the APPLGRID code~\cite{Carli:2010rw} interfaced to the MCFM program~\cite{Campbell:2010ff-1,Campbell:2010ff-2} for fast calculation of the differential cross sections at NLO in QCD and LO in electroweak (EW) couplings. Corrections to higher orders are implemented using a $K$-factor technique, correcting on a bin-by-bin basis from NLO to NNLO in QCD and from LO to NLO for the EW contribution \cite{ref2628of1612.03106-1,ref2628of1612.03106-2}.
 
Predictions for $W$~+~jets and $Z$~+~jets production are obtained similarly to the $W$ and $Z$ predictions to NLO in QCD and LO in EW couplings by using the APPLGRID code interfaced to the MCFM program. Higher-order corrections are implemented as $K$-factors.
For the $W$~+~jets data, the $N_\text{jetti}$ program~\cite{WjetNNLO} is used to calculate and implement corrections to NNLO in QCD, while the non-perturbative hadronisation and underlying event QCD corrections are computed using the \SHERPA{} v.2.2.1 MC simulation\footnote{These non-perturbative corrections account for the collective effect of using jets clustered from showered partons to jets clustered from hadrons with multi-parton-interactions and underlying event simulation enabled \cite{Dasgupta:2007wa,Buttar:2008jx}.} \cite{Bothmann:2019yzt,Schonherr:2017qcj,Kallweit:2015dum}.
The bin-by-bin $K$-factors are derived as the ratio of the NNLO to the NLO calculation from $N_\text{jetti}$ with the same fiducial selection as the $W$~+~jets data, multiplied by the non-perturbative correction.
The renormalisation and factorisation scales are set to $\muR = \muF =\sqrt{m_W^2+\Sigma (\pt^j)^2}$, where $m_W$ is the mass of the $W$ boson and the second term in the square root is the scalar sum of the squared transverse momenta of the jets.
More details about the predictions are given in the respective ATLAS publication~\cite{STDM-2016-14}.
In addition to these predictions, NLO EW corrections inclusive of QED radiation effects are computed using \SHERPA{} v.2.2.10 by the authors of Refs.~\cite{Bothmann:2019yzt,Schonherr:2017qcj,Kallweit:2015dum} and applied as additional bin-by-bin multiplicative $K$-factors.
 
Predictions for $Z$~+~jets production to NNLO in QCD and LO in EW couplings are calculated by the authors of Ref.~\cite{ZjetNNLO}, and the $K$-factor is calculated as the ratio of  NNLO to NLO predictions.
The renormalisation and factorisation scales are set to $\muR = \muF = \frac{1}{2}(\sum p_{\text{T},\mathrm{partons}} + \sqrt{m_{\ell\ell}^2 + p_{\text{T},\ell\ell}^2})$ where $m_{\ell\ell}$ is the electron-pair invariant mass, $p_{\text{T},\ell\ell}$ is the transverse momentum of the electron pair and $\Sigma p_{\text{T},\mathrm{partons}}$ is the sum of the transverse momenta of the outgoing partons.
Corrections for QED radiation effects and non-perturbative QCD corrections are each calculated using the \SHERPA{} v.1.4.5 MC simulation, as discussed in the publication describing the ATLAS measurement~\cite{STDM-2016-11}, and each provided as a set of bin-by-bin multiplicative $K$-factors.
Corrections for NLO EW effects excluding QED radiation are computed using \SHERPA{} v.2.2.10 and applied as additional bin-by-bin $K$-factors.
The $K$-factors for both $W$~+~jets and $Z$~+~jets production are typically within $10\%$ of unity, except for the NLO EW corrections for the $W$~+~jets predictions, which are as large as 20\% at high \WPT.
 
The DGLAP evolution equations of QCD yield the proton PDFs
at any value of $Q^2$ given that they are parameterised
as functions of $x$ at an initial scale $Q^2_0$.
In this analysis, the initial scale is
chosen to be $Q^2_0 = 1.9$~\GeV$^2$ such that it is below the
charm-mass matching scale, $\mu_{c}^{2}$, which is set equal to the charm mass, $\mu_{c} = m_{c}$.
The heavy-quark masses are set to their pole masses as determined by a combined analysis of HERA data on inclusive and heavy-flavour DIS processes \cite{herapdf20,H1:2018flt}, $m_c=1.43$~\GeV{} and $m_b=4.5$~\GeV, and the strong coupling constant is fixed to
$\alphas(m_Z) =  0.118$.
These choices follow those of the HERAPDF2.0 fit \cite{herapdf20}.
% End of text imported from the .//sections/Theory.tex input file
 
% The next lines are included from the .//sections/Method.tex input file
 
The quark distributions at the initial scale are assumed to behave according to the following parameterisation also used by the HERAPDF2.0 and ATLASepWZ16 fits \cite{herapdf20,STDM-2012-20}
\begin{equation}
xq_i(x) = A_i x^{B_i} (1-x)^{C_i} P_i(x)\, ,
\label{eqn:pdf}
\end{equation}
where $P_i(x)= (1 + D_i x +E_i x^2) \mathrm{e}^{F_ix}$.
The parameterised quark distributions, $xq_i$, are chosen to be
the valence-quark distributions ($xu_v,~xd_v$) and the
light-antiquark distributions
($x\bar{u},~x\bar{d},~x\bar{s}$). The gluon distribution
is parameterised with the more flexible form
\begin{equation}\nonumber
xg(x) = A_g x^{B_g} (1-x)^{C_g}P_g(x)  - A'_g x^{B'_g} (1-x)^{C'_g},
\end{equation}
where $C'_g$ is fixed to a value of $25$ to suppress negative contributions from the primed term at high $x$, as in Ref.~\cite{0901.0002}.
The parameters $A_{u_v}$ and $A_{d_v}$ are constrained using the quark counting rules, and $A_g$ is constrained using the momentum sum rule.
The normalisation and slope parameters, $A$ and $B$,
of the $\bar{u}$ and $\bar{d}$ PDFs are set equal such that
$x\bar{u} = x\bar{d}$ as  $x\to 0$. The strange PDF
$x\bar{s}$ is parameterised as in \cref{eqn:pdf},
with $P_{\bar{s}}=1$ and $B_{\bar{s}} = B_{\bar{d}}$, leaving
two free parameters for the strange PDF, $A_{\sbar}$ and $C_{\sbar}$.
It is assumed that $xs=x\sbar$ as the data used are not sufficient to distinguish between the two.
\par
The $D,E$ and $F$ terms in the expression $P_i(x)$ are used only if required
by the data, following the procedure described
in Ref.~\cite{herapdf20}.
For the ATLASepWZVjet20 fit, this results in the usage of
two additional parameters: $E_{u_v}$ and $D_{g}$.  In total, 16 free parameters are used in the central fit.\par
The level of agreement of the data with the predictions from a PDF parameterisation is quantified with a $\chi^2$. The definition of the $\chi^2$ without statistical correlations between data points is as follows~\cite{herapdf20,STDM-2012-20}
\begin{equation} \label{eqn:chi2-2}
\chi^2 = \sum_{i} \frac{[D_i - T_i(1-\sum\limits_{j}\gamma_{ij}b_{j})]^2}{\delta^{2}_{i,\text{uncor}}T_i^2 + \delta^{2}_{i,\text{stat}}D_i T_i}
+ \sum_{j} b_j^2
+ \sum_{i} \log \frac{\delta^{2}_{i,\text{uncor}}T_i^2 + \delta^{2}_{i,\text{stat}}D_i T_i}{\delta^{2}_{i,\text{uncor}}D_i^2 + \delta^{2}_{i,\text{stat}}D_i^2}\, ,
\end{equation}
where $D_i$ represent the measured data, $T_i$ represent the corresponding theoretical prediction, $\delta_{i,\text{uncor}}$ and $\delta_{i,\text{stat}}$ are the uncorrelated systematic and statistical uncertainties in $D_i$, and correlated systematic uncertainties, described by $\gamma_{ij}$, are accounted for using the nuisance parameters $b_j$. The summation over $i$ runs over all data points and the summation over $j$ runs over all sources of correlated systematic uncertainties. For each data set, the first term gives the \emph{partial} $\chi^2$ and the second term gives the \emph{correlated} $\chi^2$. The third term is a bias correction term arising from the transition of the likelihood to $\chi^2$ when the scaling of errors is applied, referred to as the \emph{log penalty}.
For the $W$ + jets data, the bin-to-bin statistical correlations are significant in contrast to the other data sets and incorporated into the $\chi^2$ definition as follows
\begin{equation} \label{eqn:chi2-1}
\begin{aligned}
\chi^2 = &\sum_{ik}\left(D_i - T_i(1-\sum_{j}\gamma_{ij}b_{j})\right)C^{-1}_{\text{stat}, ik}(D_i,D_k)\left(D_k - T_k(1-\sum_{j}\gamma_{kj}b_{j})\right)\\
&+ \sum_{j} b_j^2 \\
&+ \sum_{i} \log \frac{\delta^{2}_{i,\text{uncor}}T_i^2 + \delta^{2}_{i,\text{stat}}D_i T_i}{\delta^{2}_{i,\text{uncor}}D_i^2 + \delta^{2}_{i,\text{stat}}D_i^2}\, ,
\end{aligned}
\end{equation}
in which the first term of \cref{eqn:chi2-2} has been replaced with one which takes into account the diagonal and off-diagonal elements of the data statistical covariance matrix between bins $i$ and $k$, $C_{\text{stat}, ik}$.

% End of text imported from the .//sections/Method.tex input file
 
% The next lines are included from the .//sections/Results.tex input file
\section{Results}
 
In this section, the ATLASepWZVjet20 PDF set is presented and compared with an equivalent fit performed without the $V$~+~jets data, where the latter is named the ATLASepWZ20 PDF set.
These PDFs differ from the ATLASepWZ16 analysis by an additional parameter, $D_g$, a tighter selection criterion of $Q^2 > 10~\GeV^2$ and the use of ATLAS 7~\TeV{} $W$ and $Z$ data in which the electron and muon channels are not combined. The result is very similar except for a larger total uncertainty resulting from the use of more parameterisation variations.
It was verified that the use of 7~\TeV{} $W$ and $Z$ data with the electron and muon channels combined provides a fit with very similar central values and uncertainties.
Rather than being intended to supersede the ATLASepWZ16 PDF, the ATLASepWZ20 fit is provided to allow a more meaningful comparison with the ATLASepWZVjet20 fit by having a PDF set that differs only in the addition of the $V$~+~jets data.
 
% The next lines are included from the .//sections/Results/DataFits.tex input file
\subsection{Goodness of fit and parton distributions}\label{subsec:DataFits}
 
\Cref{fig:DataPlots1,fig:DataPlots2,fig:DataPlots3} show a comparison of the $W$~+~jets and $Z$~+~jets differential cross-section measurements with the predictions of the ATLASepWZ20 and ATLASepWZVjet20 fits.  Adding the $V$~+~jets data to the fit improves the $W$~+~jets description significantly, particularly in the $W^+$ spectrum, where agreement with data improves by approximately 20\% at high \WPT.
The difference in partial $\chi^2$ between the predictions of the ATLASepWZ20 and ATLASepWZVjet20 PDF sets for the $W$~+~jets and the $Z$~+~jets data is 32 and 7 units, respectively.
 
\begin{figure}
\centering
\subfloat[]{\includegraphics[width=.5\linewidth]{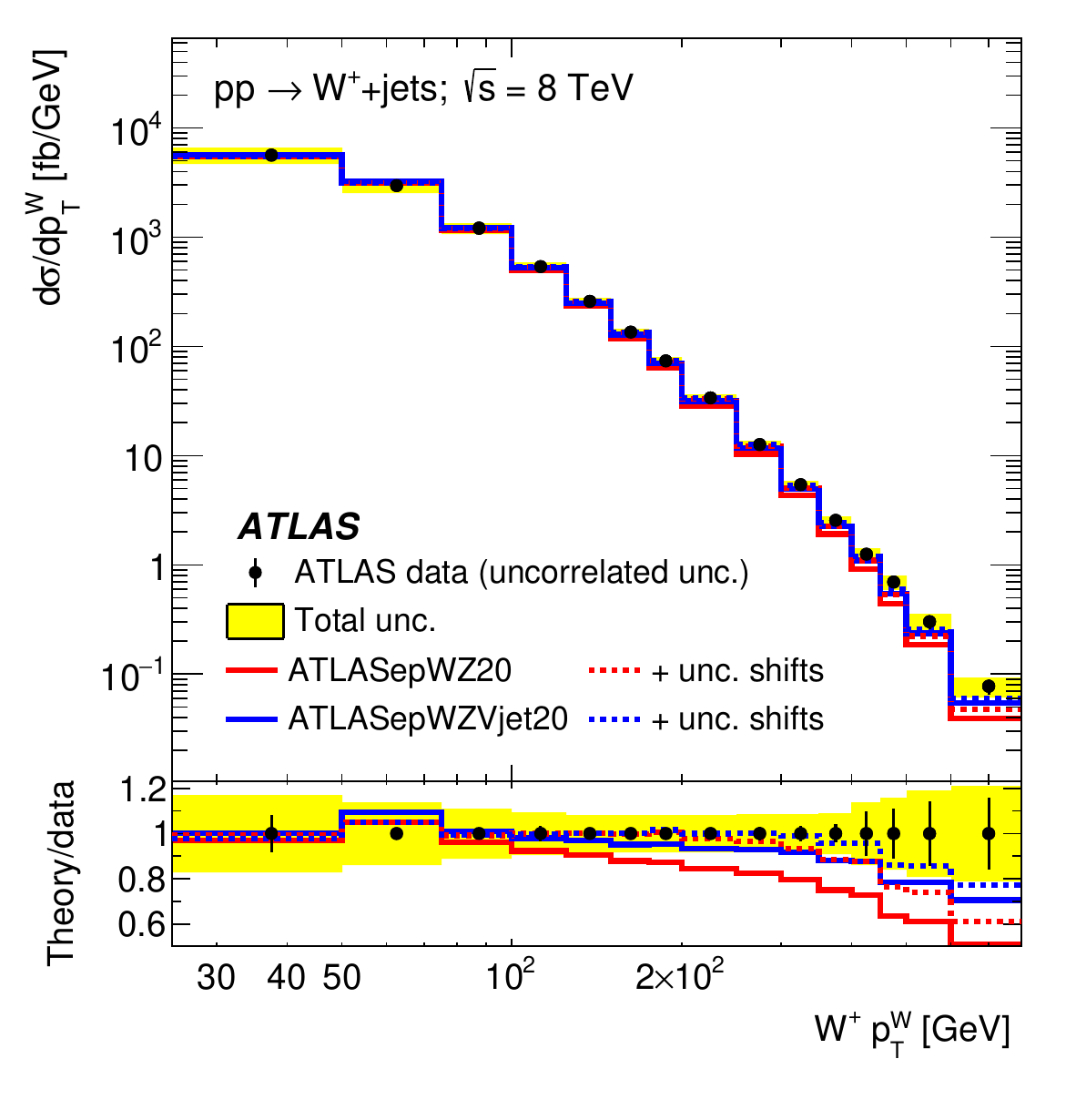}}
\subfloat[]{\includegraphics[width=.5\linewidth]{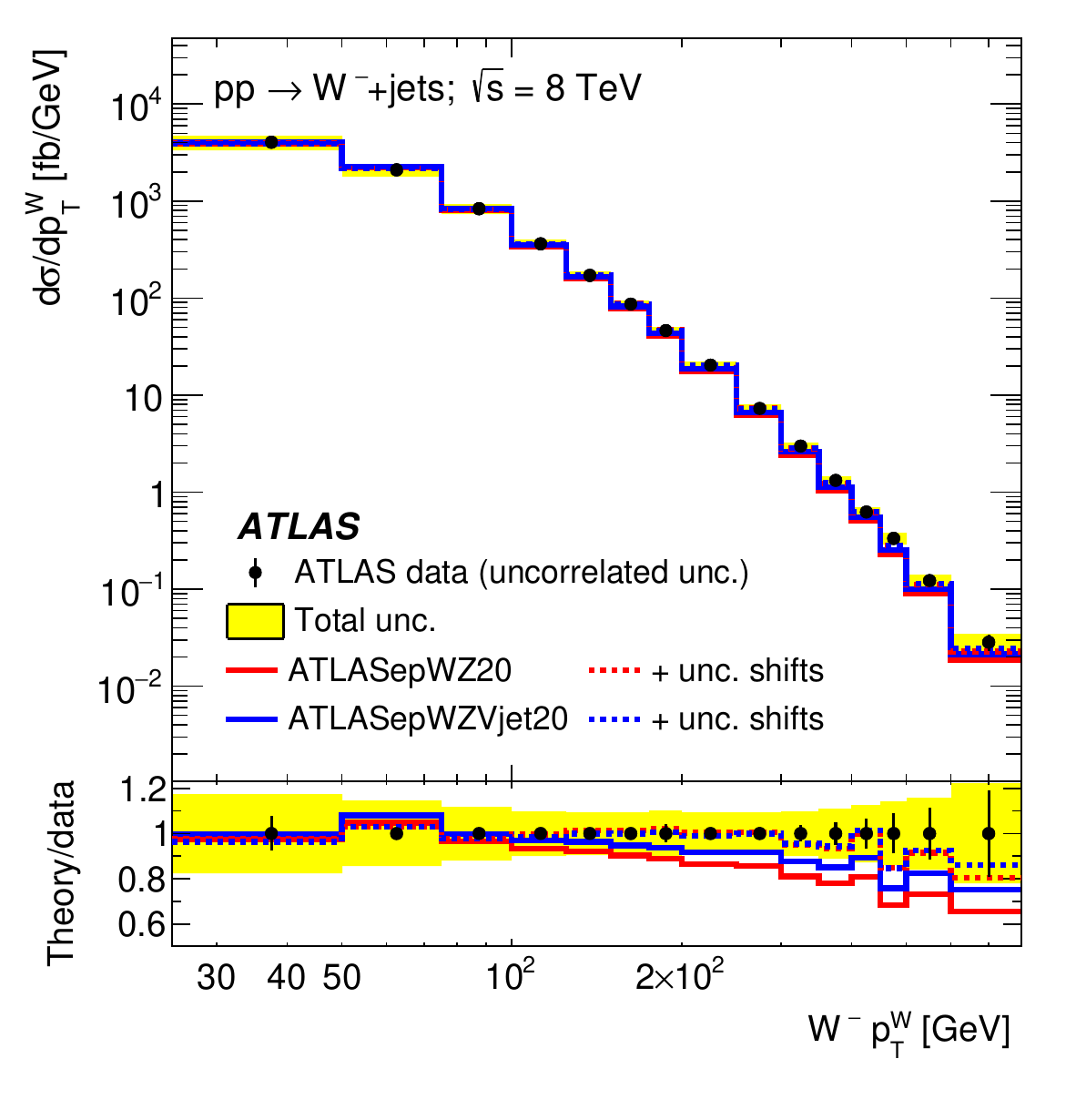}}
 
\caption{The differential cross-section measurements of (a) $W^+$~+~jets and (b) $W^-$~+~jets in Ref.~\cite{STDM-2016-14} (black points) as a function of the transverse momentum of the $W$ boson, \WPT. The bin-to-bin uncorrelated part of the data uncertainties is shown as black error bars, while the total uncertainties are shown as a yellow band. The cross sections are compared with the predictions computed with the predetermined PDFs resulting from the fits ATLASepWZ20 (red lines) and ATLASepWZVjet20 (blue lines). The solid lines show the predictions without shifts of the systematic uncertainties, while for the dashed lines the $b_j$ parameters associated with the experimental systematic uncertainties as shown in \cref{eqn:chi2-1} are allowed to vary to minimise the $\chi^2$. The ratios of each set of predictions to the data are shown in the bottom panel in each case.}
\label{fig:DataPlots1}
\end{figure}
 
\begin{figure}
\centering
\subfloat[]{\includegraphics[width=.5\linewidth]{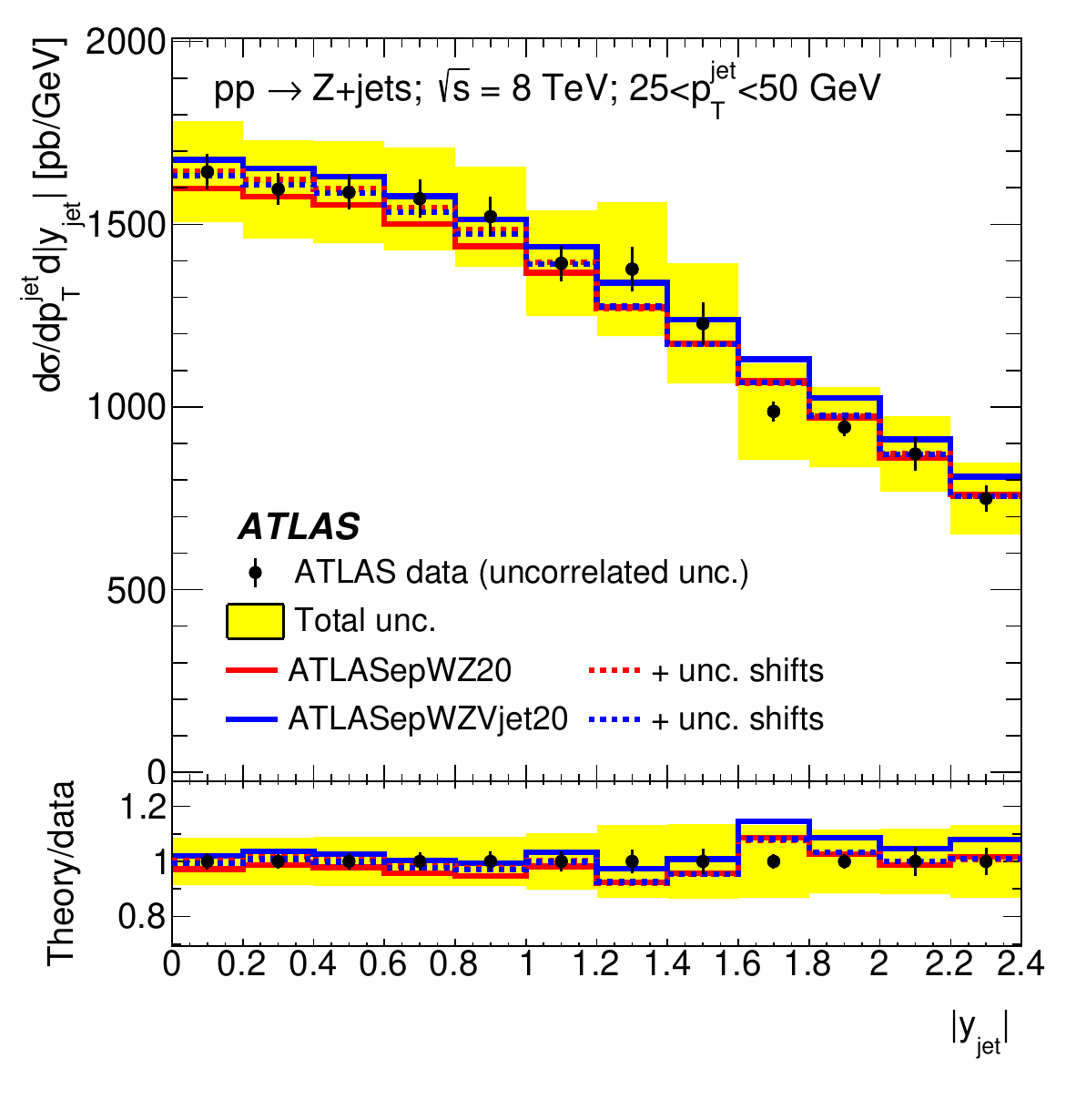}}
\subfloat[]{\includegraphics[width=.5\linewidth]{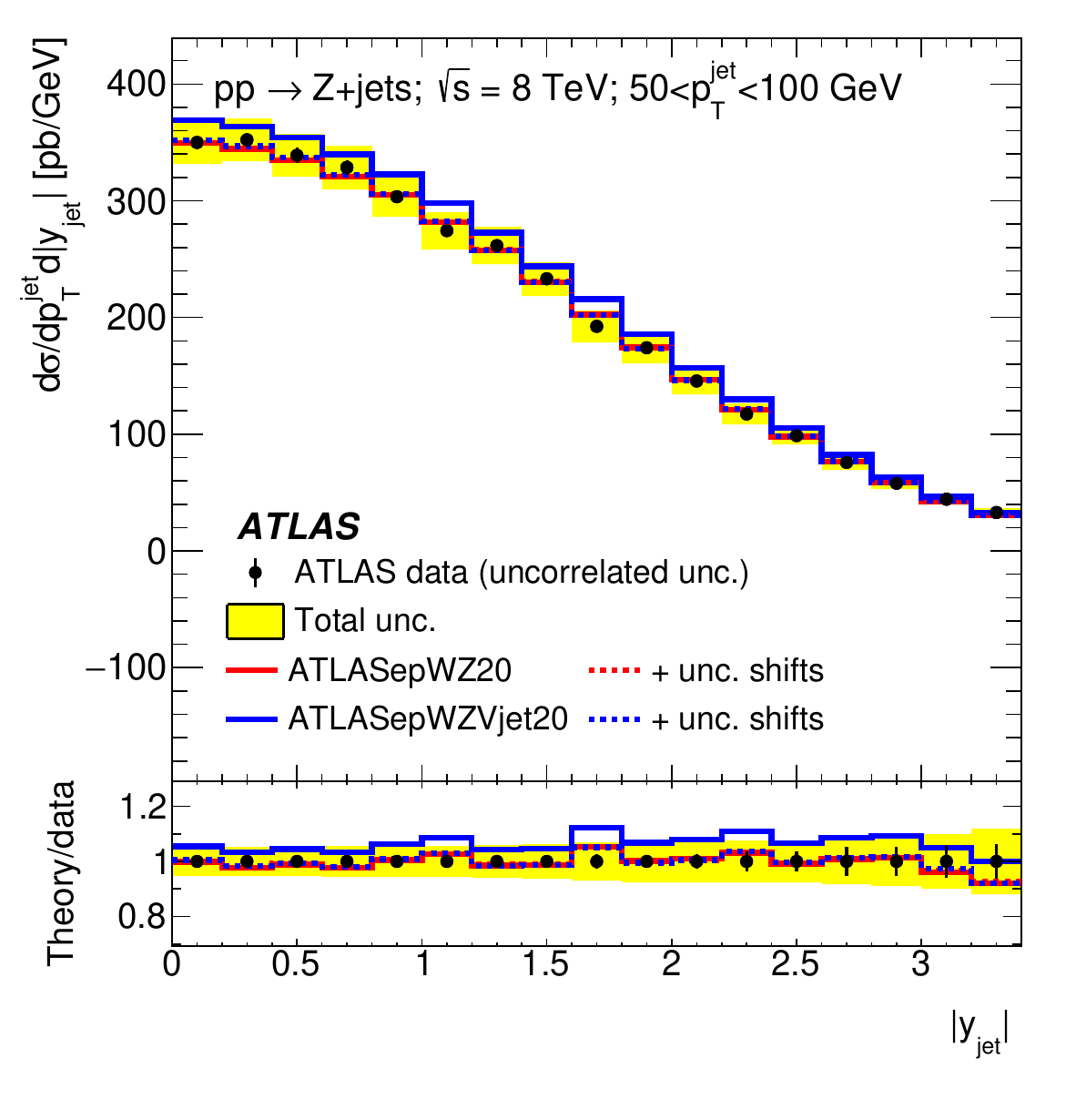}}
 
\subfloat[]{\includegraphics[width=.5\linewidth]{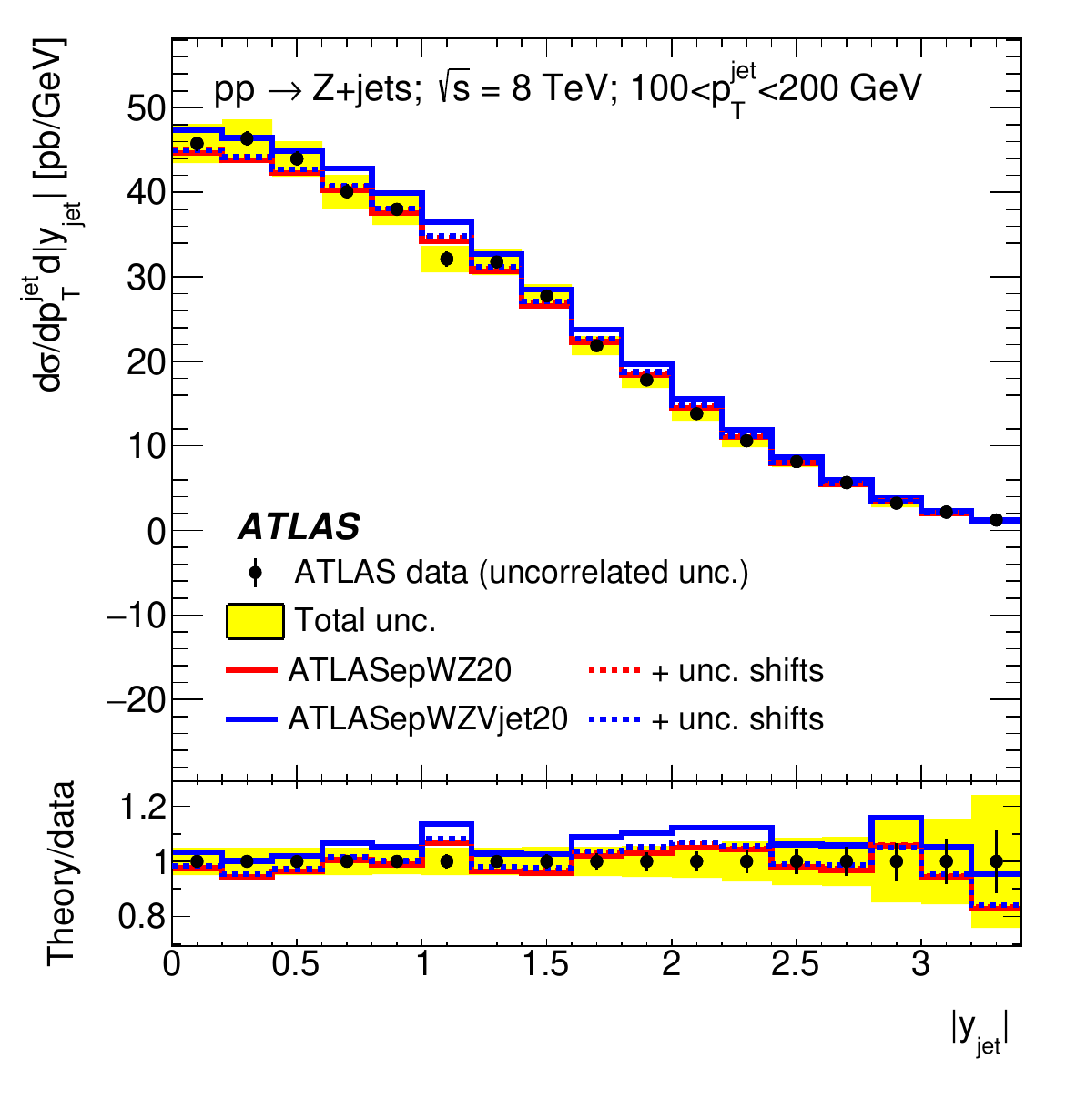}}
\subfloat[]{\includegraphics[width=.5\linewidth]{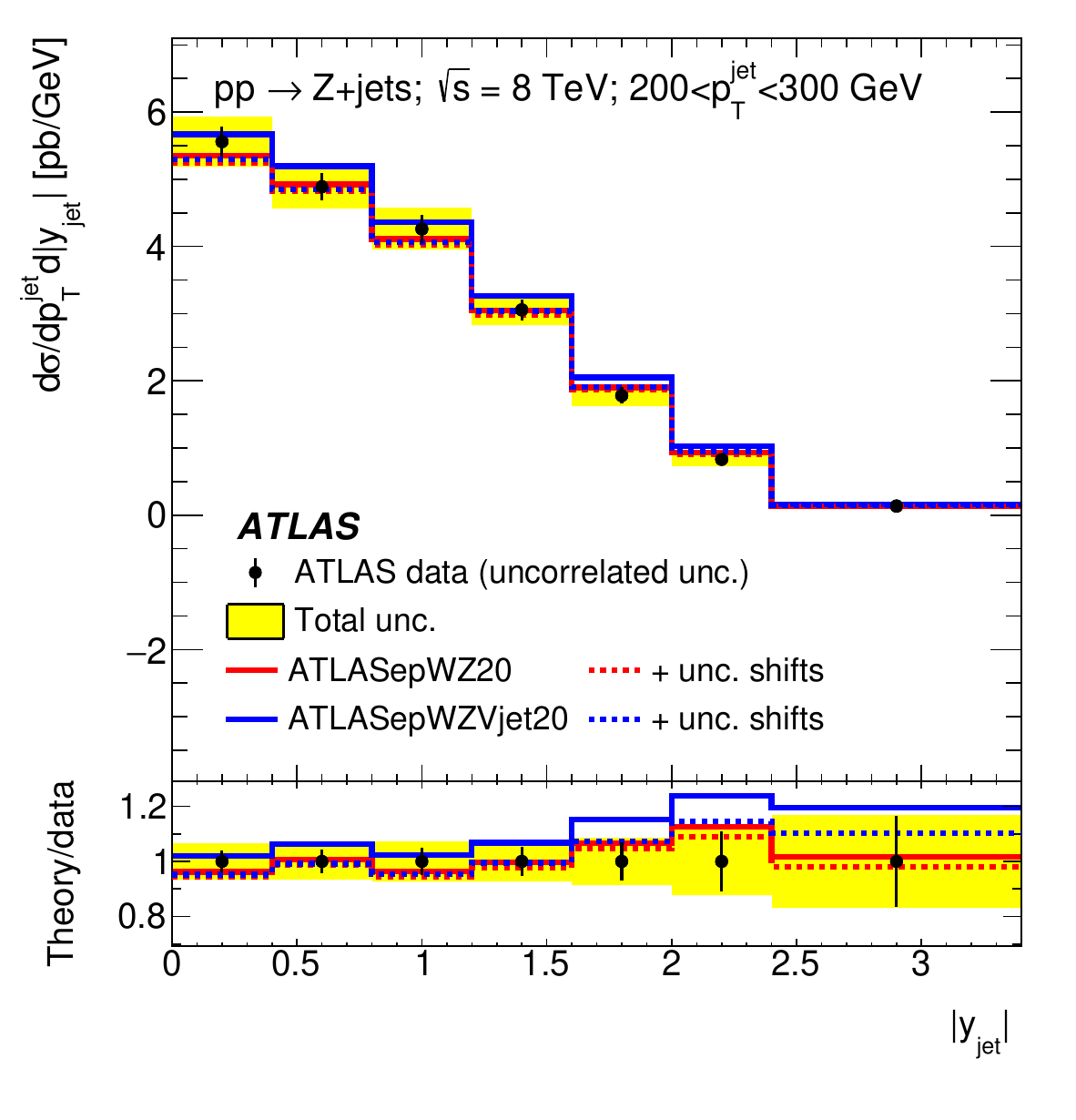}}
 
\caption{The differential cross-section measurements of $Z$~+~jets as a function of the absolute rapidity of inclusive jets, \yjets{}, in bins of (a) $25 < \ptjets < 50~\GeV$, (b) $50 < \ptjets < 100~\GeV$, (c) $100 < \ptjets < 200~\GeV$ and (d) $200 < \ptjets < 300~\GeV$, where the transverse momentum of inclusive jets is labelled \ptjets. The bin-to-bin uncorrelated part of the data uncertainties is shown as black error bars, while the total uncertainties are shown as a yellow band. The cross sections are compared to the predictions computed with the PDFs resulting from the fits ATLASepWZ20 (red lines) and ATLASepWZVjet20 (blue lines). The solid lines show the predictions without shifts of the systematic uncertainties, while for the dashed lines the shifts with fitted $b_j$ parameters as shown in \cref{eqn:chi2-1} are applied. The ratios of each set of predictions to the data are shown in the bottom panel in each case.}
 
\label{fig:DataPlots2}
\end{figure}
 
\begin{figure}
\centering
\subfloat[]{\includegraphics[width=.5\linewidth]{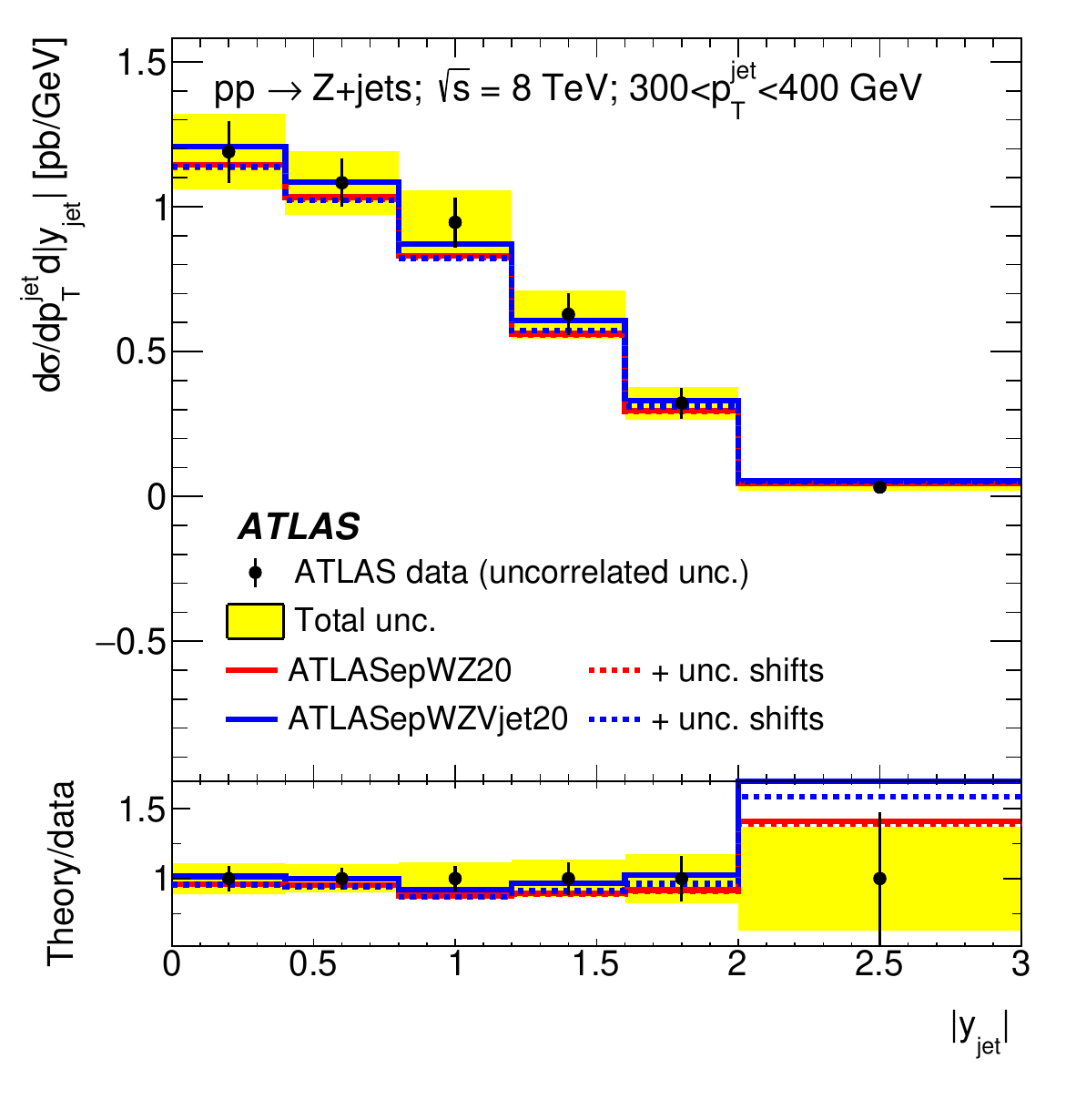}}
\subfloat[]{\includegraphics[width=.5\linewidth]{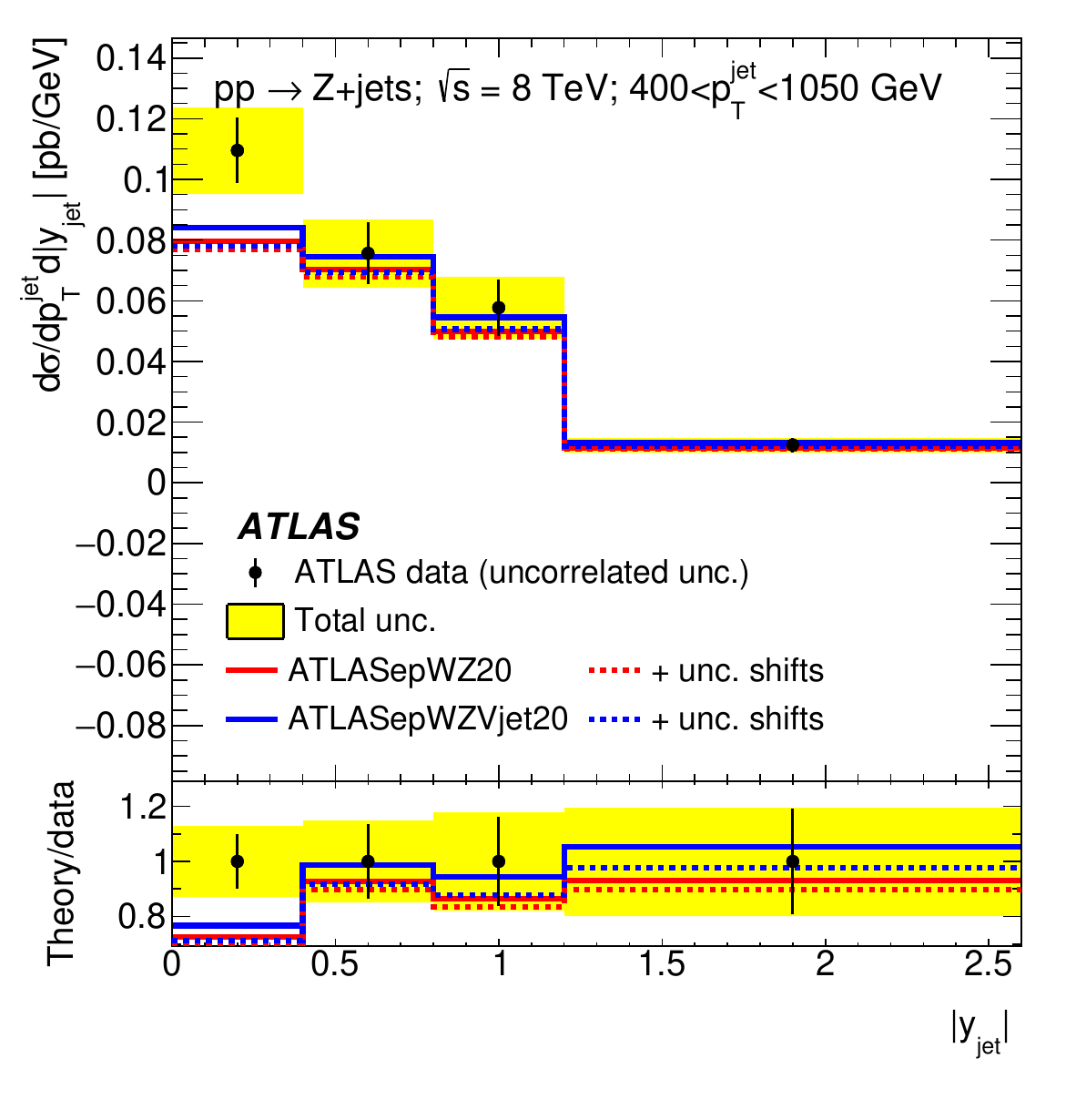}}
 
\caption{The differential cross-section measurements of $Z$~+~jets as a function of the absolute rapidity of inclusive jets, \yjets{}, in bins of (a) $300 < \ptjets < 400~\GeV$ and (b) $400 < \ptjets < 1050~\GeV$, where the transverse momentum of inclusive jets is labelled \ptjets. The bin-to-bin uncorrelated part of the data uncertainties is shown as black error bars, while the total uncertainties are shown as a yellow band. The cross sections are compared to the predictions computed with the PDFs resulting from the fits ATLASepWZ20 (red lines) and ATLASepWZVjet20 (blue lines). The solid lines show the predictions without shifts of the systematic uncertainties, while for the dashed lines the shifts with fitted $b_j$ parameters as shown in \cref{eqn:chi2-1} are applied. The ratios of each set of predictions to the data are shown in the bottom panel in each case.}
 
\label{fig:DataPlots3}
\end{figure}
 
\begin{table}[h!]
\centering
\begin{tabular}{l H c}
\hline
\hline
Fit & ATLASepWZ20 & ATLASepWZVjet20\\
\hline
Total $\chi^2/\mathrm{NDF}$ & 1316 / 1105 & 1460 / 1198\\
HERA partial $\chi^2/\mathrm{NDP}$ & 1114 / 1016 & 1132 / 1016\\
HERA correlated $\chi^2$ & 49 & 50\\
HERA log penalty $\chi^2$ & $-11$~~~~ & $-12$~~~~\\
ATLAS $W, Z$ partial $\chi^2/\mathrm{NDP}$ & 121 / 105 & 113 / 105\\
ATLAS $W$ + jets partial $\chi^2/\mathrm{NDP}$ & 0 / 0 & 25 / 30\\
ATLAS $Z$ + jets partial $\chi^2/\mathrm{NDP}$ & 0 / 0 & 82 / 63\\
ATLAS correlated $\chi^2$ & 41 & 65\\
ATLAS log penalty $\chi^2$ & ~~1 & ~~6\\
\hline
\hline
\end{tabular}
\caption{$\chi^2$ values split by partial, correlated and log penalty for each data set entering the ATLASepWZVjet20 fit. The partial component of the $\chi^2$ for each data set is shown compared with the number of data points of that data set (NDP).}
\label{tab:chi}
\end{table}
 
The total $\chi^2$ per degree of freedom ($\chi^2$/NDF) for the ATLASepWZVjet20 fit, along with the partial $\chi^2$ per data point ($\chi^2$/NDP) and correlated $\chi^2$ for each data set entering the fit, is given in \cref{tab:chi}.
The partial $\chi^2$ for the HERA and ATLAS inclusive $W$ and $Z$ data in the ATLASepWZVjet20 fit is similar to those obtained in the ATLASepWZ20 fit, not showing any tension between these data and the $V$ + jets data.
The partial $\chi^2$ of the $W$~+~jets and $Z$~+~jets data is reasonable, and neither the HERA nor ATLAS correlated $\chi^2$ is observed to increase significantly with the inclusion of this data.
 
Additional uncertainties in the PDFs are estimated and classified as either model or parameterisation uncertainties.
Model uncertainties comprise variations of the charm-quark mass ($m_c$) and bottom-quark mass ($m_b$), variations of the minimum $Q^2$ cut, $Q^2_{\mathrm{min}}$, and the starting scale at which the PDFs are parameterised, $Q_0^2$.
The variation in charm-quark mass and starting scale are performed simultaneously to fulfil the condition $Q_0^2 < m_c^2$ such that the charm PDF is calculated perturbatively.
Each of these variations follow that of the ATLASepWZ16 analysis \cite{STDM-2012-20}.
The parameterisation uncertainties are estimated through variations which include a single further parameter in the polynomial $P_i(x)$ or relaxed constraints on the low-$x$ sea quarks.
In each variation, listed with its respective total $\chi^2$ per degree of freedom in \cref{tab:ModPar}, the uncertainty is calculated as the difference between the alternative extracted PDF and the nominal PDF at each value of $x$ and $Q^2$.
Whereas the model variations are treated independently and the model uncertainty is calculated as the sum in quadrature of the variations, the parameterisation uncertainty is taken as the envelope of the parameterisation variations.
The total uncertainty is calculated as the sum in quadrature of the experimental, model and parameterisation uncertainties.
While the total uncertainty does give an estimate of the total variability of the fit, only the experimental uncertainty is interpretable similarly to a statistical standard deviation.
 
The impact of theoretical uncertainties in the $V$~+~jets predictions
on the fit results is cross-checked. Variations of the NNLO QCD
calculations are defined from the variations of
factorisation and renormalisation scales by factors of two up and down
and taking the envelope of these predictions. In the fit, the
corresponding $K$-factors are varied for the $W$~+~jets and $Z$~+~jets
prediction upward and downward both simultaneously and
individually. Each of these variations results in PDFs well within
the experimental uncertainties of the nominal ATLASepWZVjet20 set.
 
\begin{table}[h!]
\centering
\begin{tabular}{l H c}
\hline
\hline
Nominal $\chi^2$/NDF & 1316 / 1105 & 1460 / 1198\\
\hline
\multicolumn{3}{c}{Parameter variations}\\
\hline
$A_{\overline{u}} \neq A_{\overline{d}}$ & 1313 / 1104 & 1458 / 1197\\
$A_{\overline{u}} \neq A_{\overline{d}}\, \&\, B_{\overline{u}} \neq B_{\overline{d}}$ & 1313 / 1103 & 1454 / 1196\\
$B_{\overline{s}} \neq B_{\overline{d}}$ & 1313 / 1104 & 1459 / 1197\\
$B_{\overline{u}} \neq B_{\overline{d}}$ & 1313 / 1104 & 1459 / 1197\\
$D_{\overline{d}}$ & 1316 / 1104 & 1459 / 1197\\
$D_{d_{v}}$ & 1316 / 1104 & 1460 / 1197\\
$D_{\overline{s}}$ & 1316 / 1104 & 1460 / 1197\\
$D_{u_{v}}$ & 1312 / 1104 & 1457 / 1197\\
$E_{\overline{d}}$ & 1315 / 1104 & 1459 / 1197\\
$E_{\overline{s}}$ & 1313 / 1104 & 1460 / 1197\\
$E_{\overline{u}}$ & 1315 / 1104 & 1459 / 1197\\
$F_{d_{v}}$ & 1315 / 1104 & 1460 / 1197\\
$F_{\overline{s}}$ & 1310 / 1104 & 1460 / 1197\\
$F_{\overline{u}}$ & 1313 / 1104 & 1458 / 1197\\
$F_{u_{v}}$ & 1313 / 1104 & 1456 / 1197\\
\hline
\multicolumn{3}{c}{Model variations}\\
\hline
$Q^{2}_{\mathrm{min}} = 12.5~\GeV^{2}$ & 1245 / 1056 & 1393 / 1149\\
$Q^{2}_{\mathrm{min}} = 7.5~\GeV^{2}$ & 1374 / 1145 & 1529 / 1238\\
$Q^{2}_{0} = 2.2~\GeV^{2}$ and $m_{c} = 1.49~\GeV$ & 1315 / 1105 & 1465 / 1198\\
$Q^{2}_{0} = 1.6~\GeV^{2}$ and $m_{c} = 1.37~\GeV$ & 1318 / 1105 & 1458 / 1198\\
$\alphas(m_{Z}) = 0.120$ & 1317 / 1105 & 1463 / 1198\\
$\alphas(m_{Z}) = 0.116$ & 1314 / 1105 & 1458 / 1198\\
$m_{b} = 4.75~\GeV$ & 1316 / 1105 & 1461 / 1198\\
$m_{b} = 4.25~\GeV$ & 1315 / 1105 & 1458 / 1198\\
\hline
\hline
\end{tabular}
\caption{Total $\chi^2$/NDF for each parameterisation and model variation contributing to the parameterisation and model uncertainties, respectively, of the ATLASepWZVjet20 fit. Where a $D$, $E$ or $F$ parameter is referred to, this means that the respective parameter is not constrained to zero in that variation. Where two $A$ or $B$ parameters are referred to in an inequality, this means that the respective two parameters are free to vary independently of each other in a fit.}
\label{tab:ModPar}
\end{table}
 
% End of text imported from the .//sections/Results/DataFits.tex input file
% The next lines are included from the .//sections/Results/ResultingPDFs.tex input file

\Cref{fig:VJetsPDFs_1,fig:VJetsPDFs_2} shows the ATLASepWZVjet20 PDFs overlaid with the ATLASepWZ20 PDFs, each evaluated at the starting scale $Q_0^2$, for comparison.
The experimental and total uncertainties are displayed separately in each case.
The ATLASepWZVjet20 \xdbar{} distribution is notably higher in the range $x \gtrsim 0.02$ compared to the ATLASepWZ20 fit.
In contrast, the \xsbar{} distribution of the ATLASepWZVjet20 fit in the same region is lower.
Together, the differences observed between the ATLASepWZ20 and ATLASepWZVjet20 PDFs allow for an increase in the $W^{+}$ cross section, as depicted in \Cref{fig:DataPlots1}, while keeping the total down-type sea $x\bar{D} = \xdbar + \xsbar$ distribution almost unchanged up to $x \sim 0.1$.
Additionally, the $d_v$ distribution is reduced in the ATLASepWZVjet20 fit at high $x$ and increased at low $x$, compensating for the changes in the other PDFs and resulting in an $xD = xd_v + \xdbar + \xsbar$ distribution which is similar at high $x$.
The up-type quark and gluon distributions are similar between the two fits.
 
\begin{figure}
\centering
\subfloat[]{\includegraphics[width=.5\linewidth]{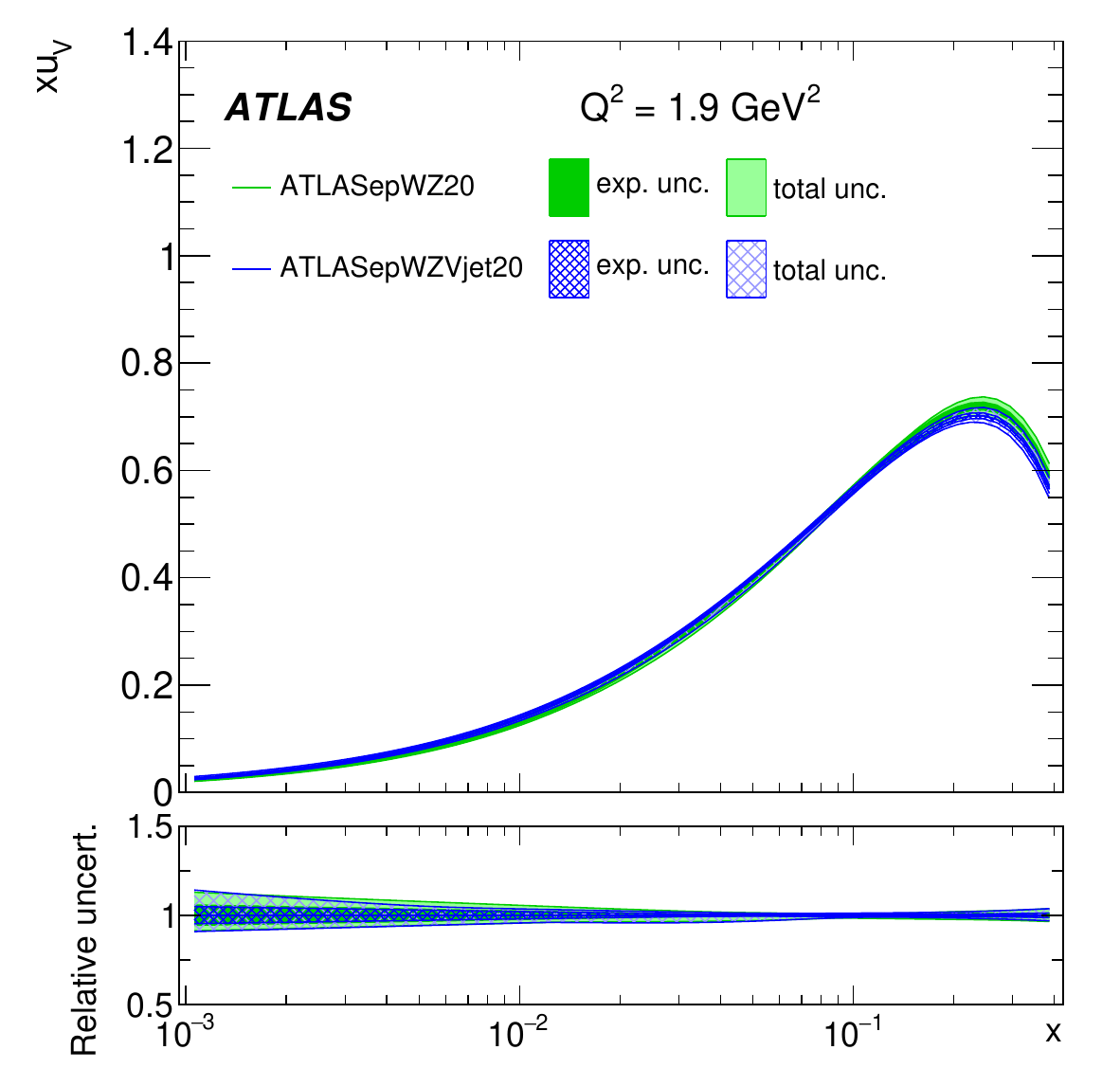}}
\subfloat[]{\includegraphics[width=.5\linewidth]{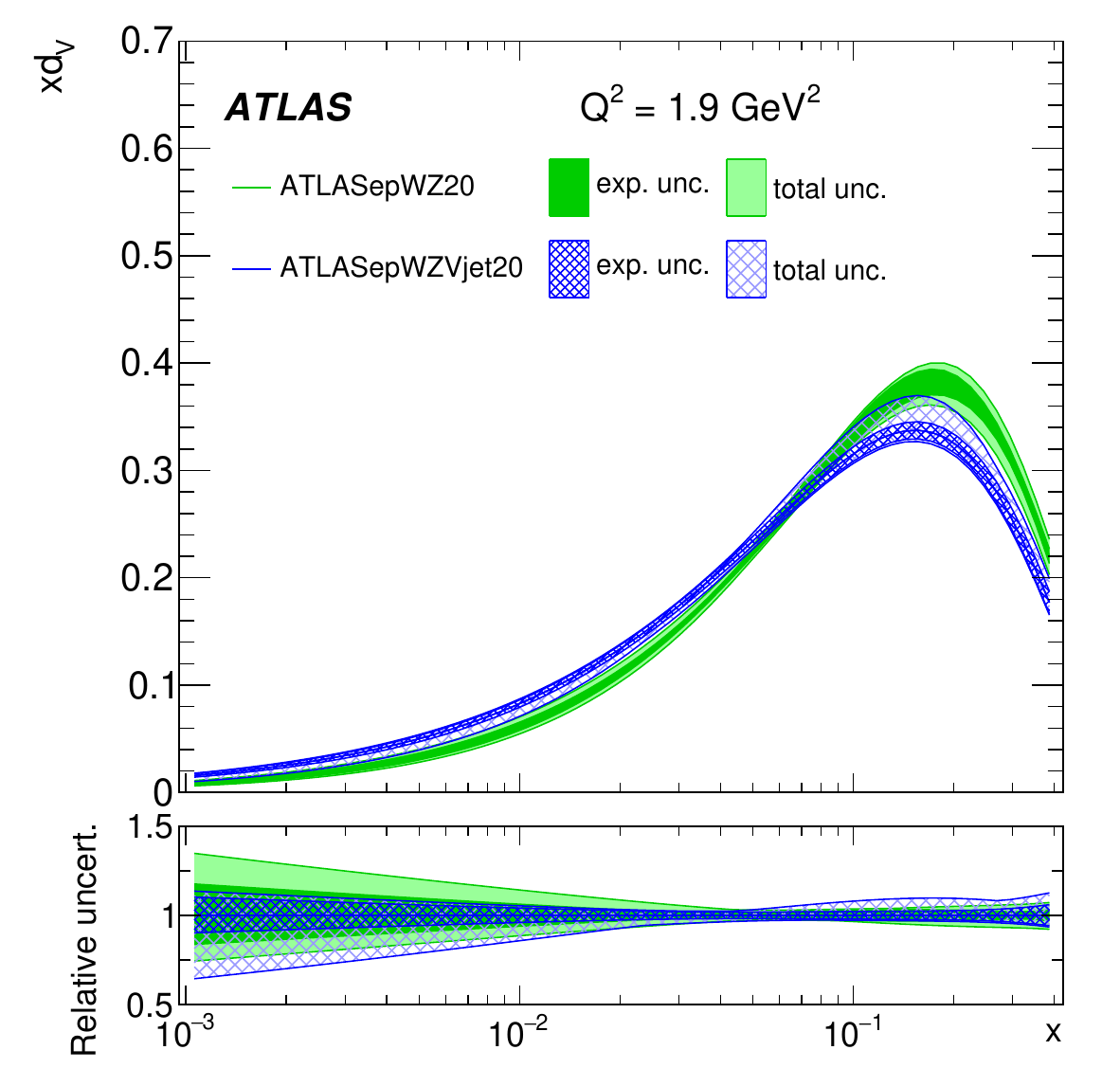}}
 
\subfloat[]{\includegraphics[width=.5\linewidth]{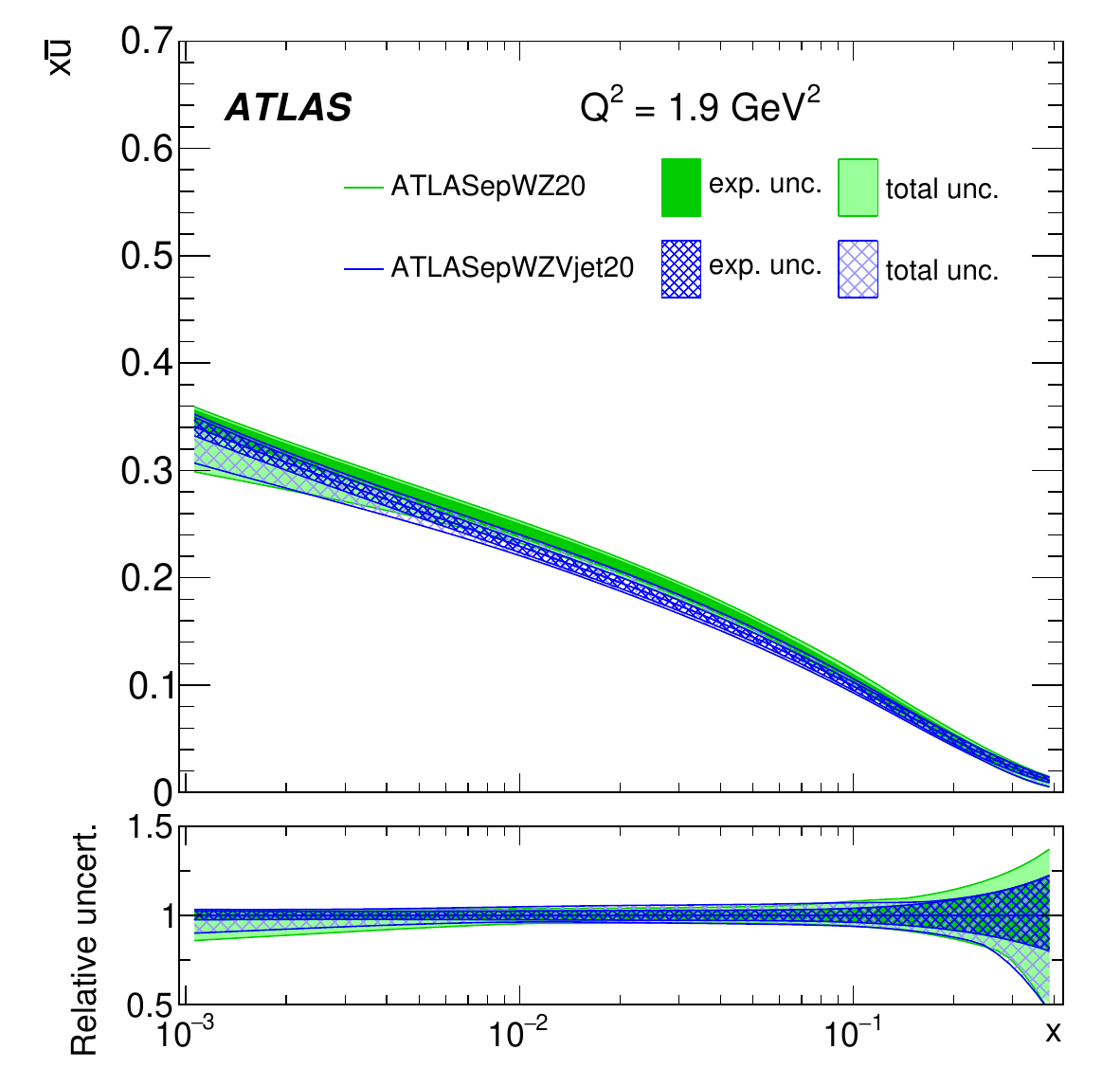}}
\subfloat[]{\includegraphics[width=.5\linewidth]{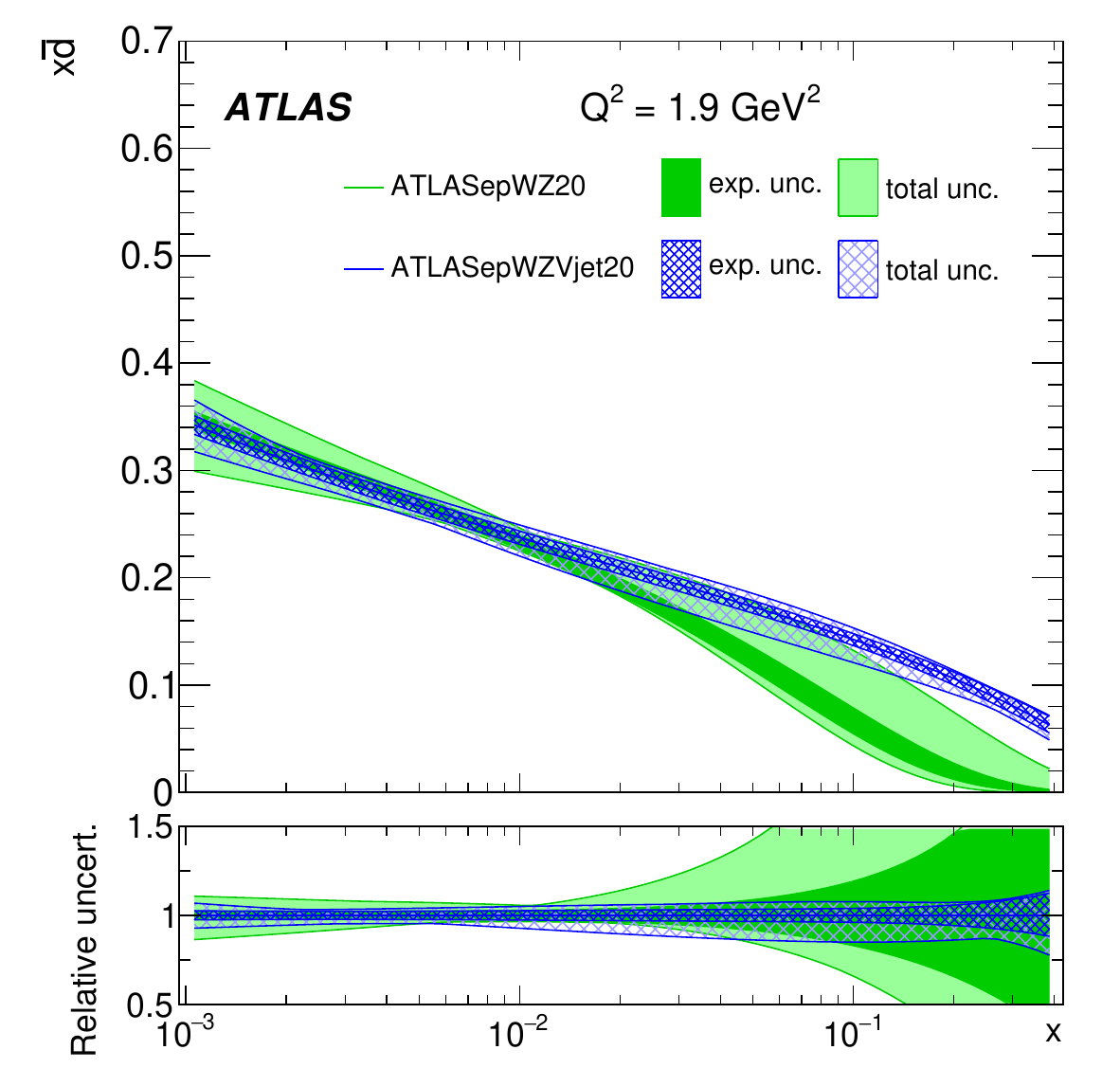}}
\caption{PDFs multiplied by Bjorken $x$ at the scale $Q^2 = 1.9~\GeV^2$ as a function of Bjorken $x$ obtained for the (a)-(b) valence quarks and (c)-(d) up and down sea quarks when fitting $W$~+~jets, $Z$~+~jets, inclusive $W$ and $Z$, and HERA data (ATLASepWZVjet20, blue bands), compared with a similar fit without $W$~+~jets or $Z$~+~jets data (ATLASepWZ20, green bands). Inner error bands indicate the experimental uncertainty, while outer error bands indicate the total uncertainty, including parameterisation and model uncertainties. The relative uncertainties around the nominal value of each PDF centred on 1 is displayed in the bottom panel in each case.}
\label{fig:VJetsPDFs_1}
\end{figure}
 
\begin{figure}
\centering
\subfloat[]{\includegraphics[width=.5\linewidth]{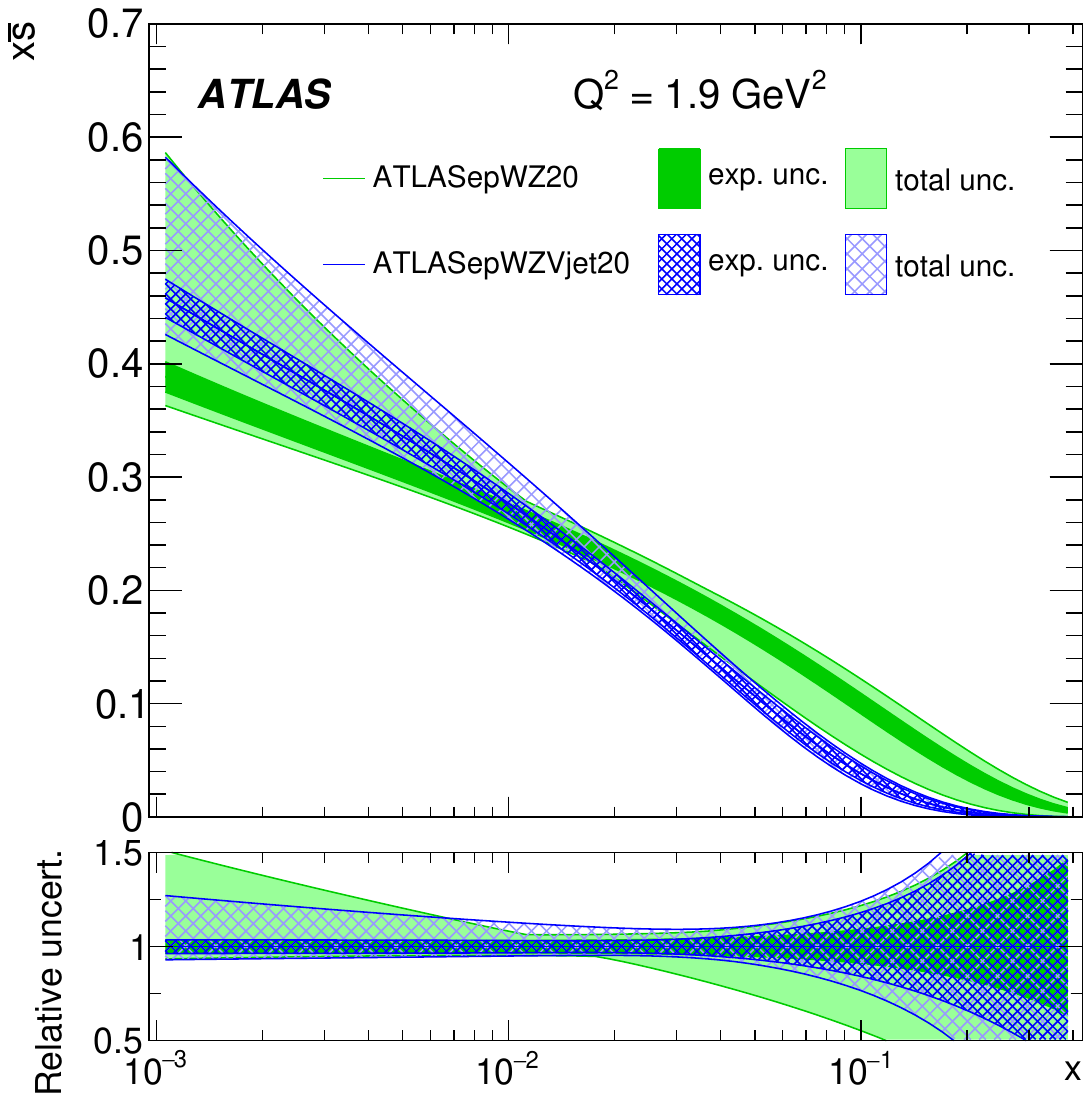}}
\subfloat[]{\includegraphics[width=.5\linewidth]{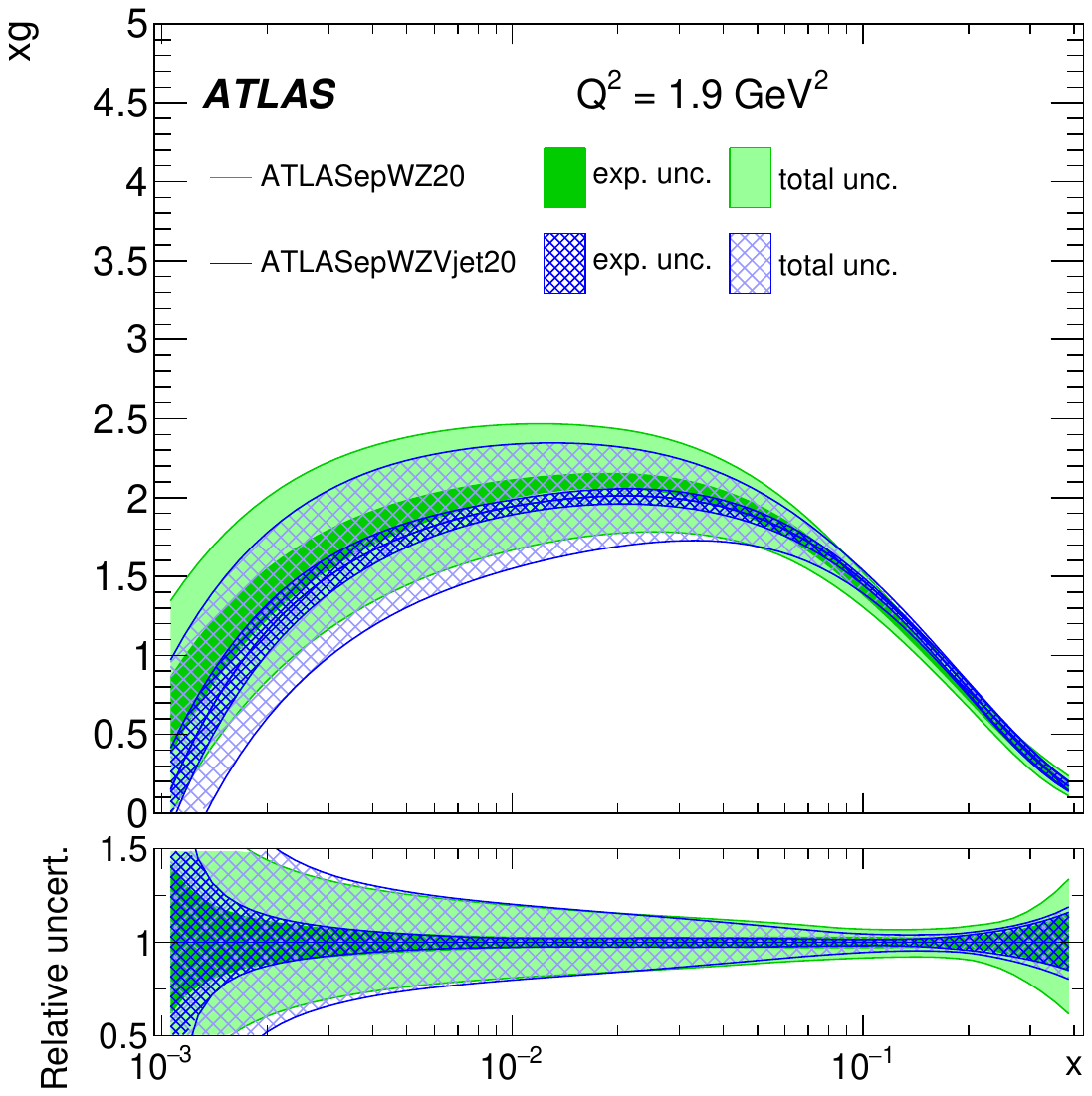}}
 
\subfloat[]{\includegraphics[width=.5\linewidth]{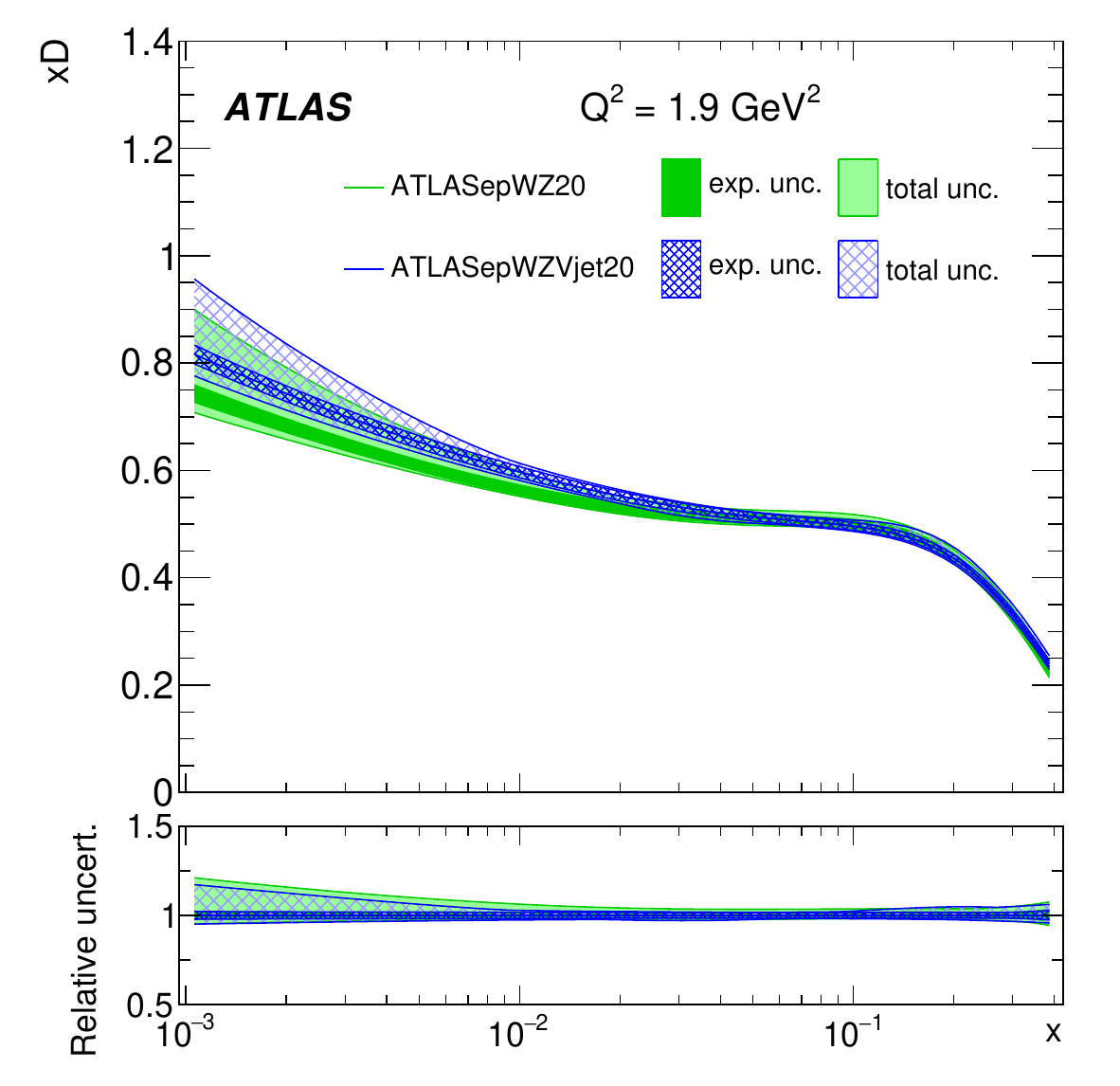}}
\subfloat[]{\includegraphics[width=.5\linewidth]{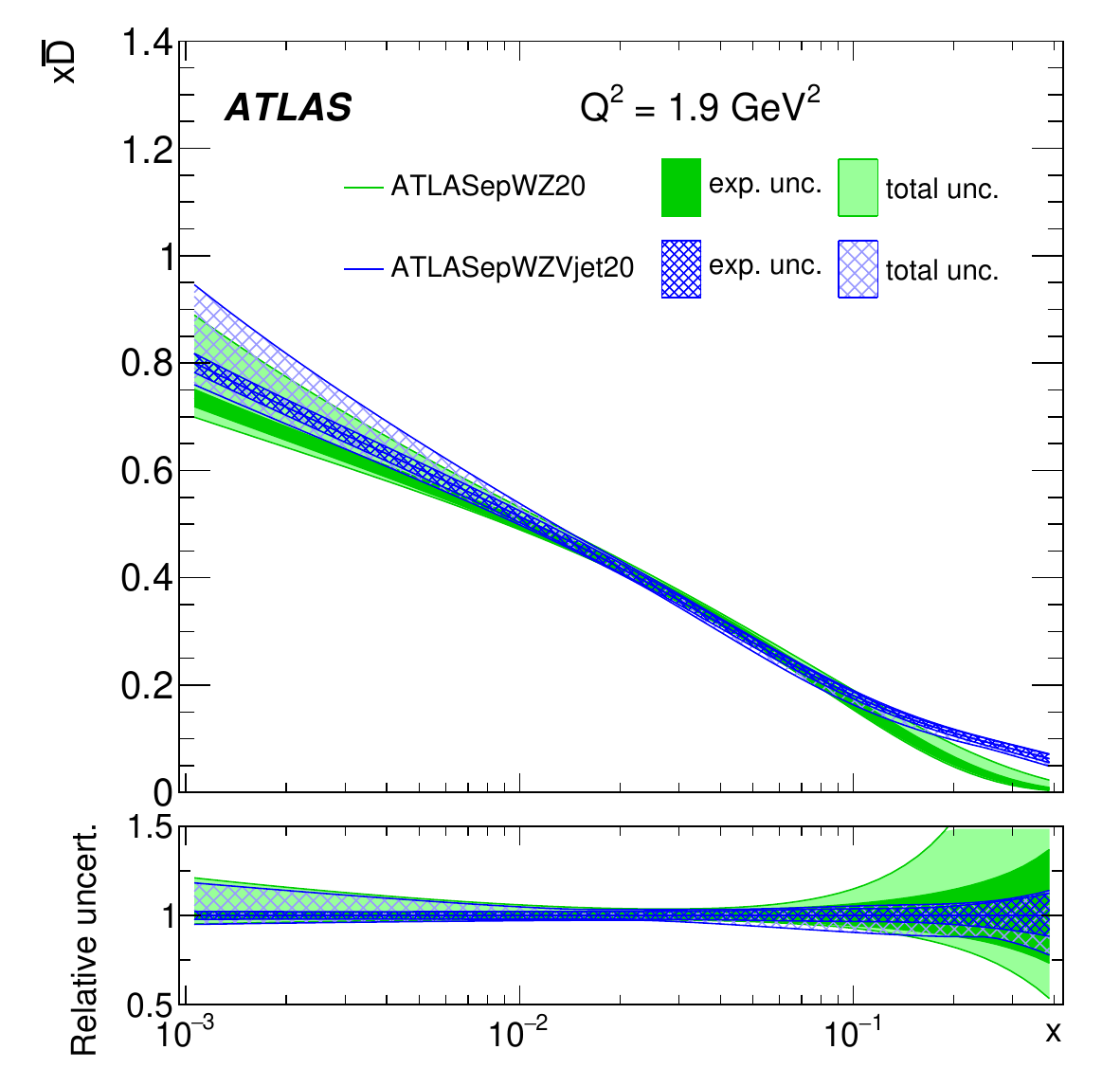}}
\caption{PDFs multiplied by Bjorken $x$ at the scale $Q^2 = 1.9~\GeV^2$ as a function of Bjorken $x$ obtained for the (a) strange sea quark, (b) gluon, (c) the total of the down-type quarks and (d) the total of the anti-down-type quarks when fitting $W$~+~jets, $Z$~+~jets, inclusive $W$ and $Z$, and HERA data (ATLASepWZVjet20, blue bands), compared with a similar fit without $W$~+~jets or $Z$~+~jets data (ATLASepWZ20, green bands). Inner error bands indicate the experimental uncertainty, while outer error bands indicate the total uncertainty, including parameterisation and model uncertainties. The relative uncertainties around the nominal value of each PDF centred on 1 is displayed in the bottom panel in each case.}
\label{fig:VJetsPDFs_2}
\end{figure}
% End of text imported from the .//sections/Results/ResultingPDFs.tex input file
% The next lines are included from the .//sections/Results/dbarubar.tex input file
\subsection{The high-$x$ sea-quark distributions}
 
The difference between the \xdbar{} and \xubar{} PDFs at high $x$ has been a topic of debate over the recent decades.
A measurement by the E866 Collaboration of the Drell--Yan cross-section ratios from an 800~\GeV{} proton beam incident on liquid hydrogen and deuterium targets found the proton $x(\dbar-\ubar)$ distribution to be positive at high $x$, peaking at $x(\bar{d}-\bar{u}) \sim 0.04$ at $x \sim 0.1$ \cite{E866}.
In contrast, the ATLASepWZ16 PDF set gives a negative central distribution with its lowest value at $x(\bar{d}-\bar{u}) \sim -0.035$ for $x \sim 0.1$, although the uncertainties are such that it is compatible with zero within two standard deviations.
 
The $x(\dbar - \ubar)$
distribution as function of Bjorken $x$ at $Q^2 = 1.9~\GeV{}^2$ is shown in
\Cref{fig:dbarubarWJets}, with a comparison between ATLASepWZVjet20 and ATLASepWZ20 displaying the direct effect of the $V$~+~jets data, and with the experimental, model and parameterisation uncertainties plotted separately.
The impact of the $V$ + jets data is to place significant constraints on the total uncertainty at high $x$, with an overall positive distribution of central values driven by the increase in the high-$x$ \dbar distribution, as discussed in \cref{subsec:DataFits}.
 
\begin{figure}
\centering
\subfloat[]{\includegraphics[width=.49\linewidth]{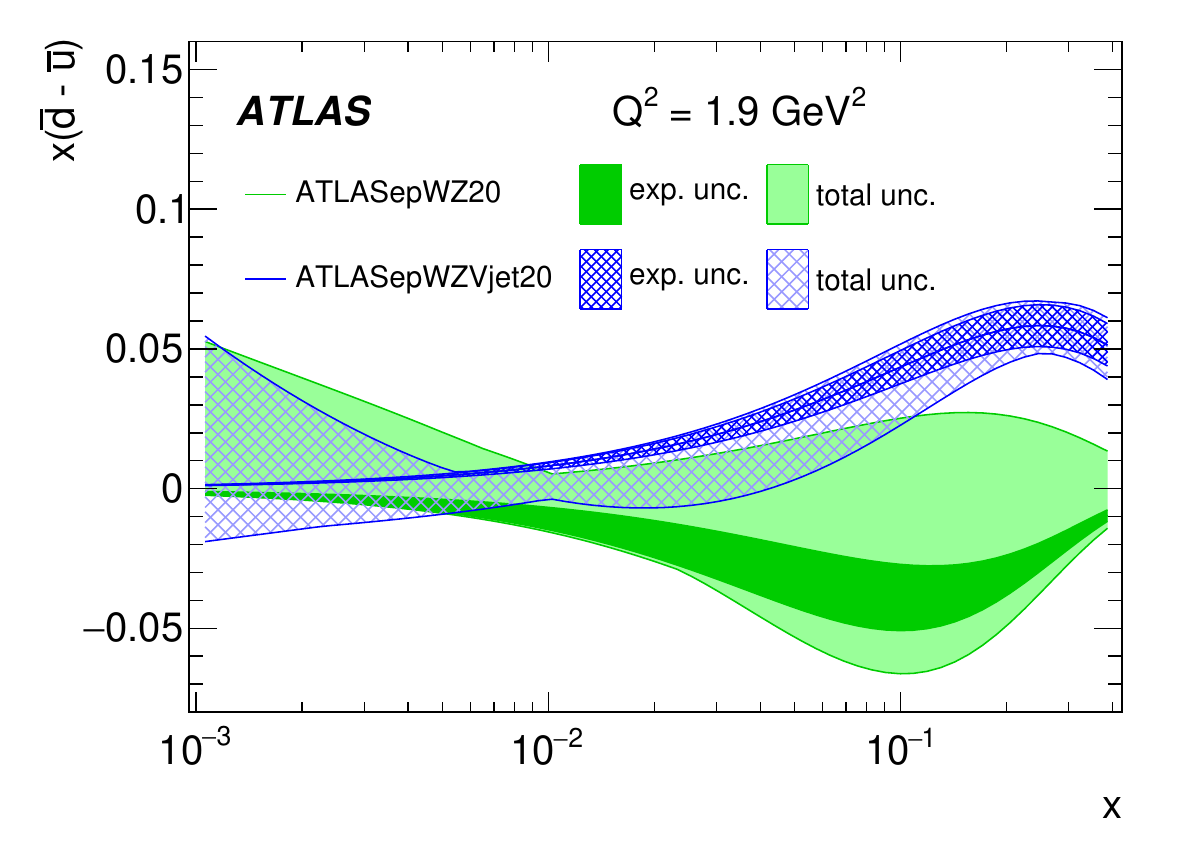}}
\subfloat[]{\includegraphics[width=.49\linewidth]{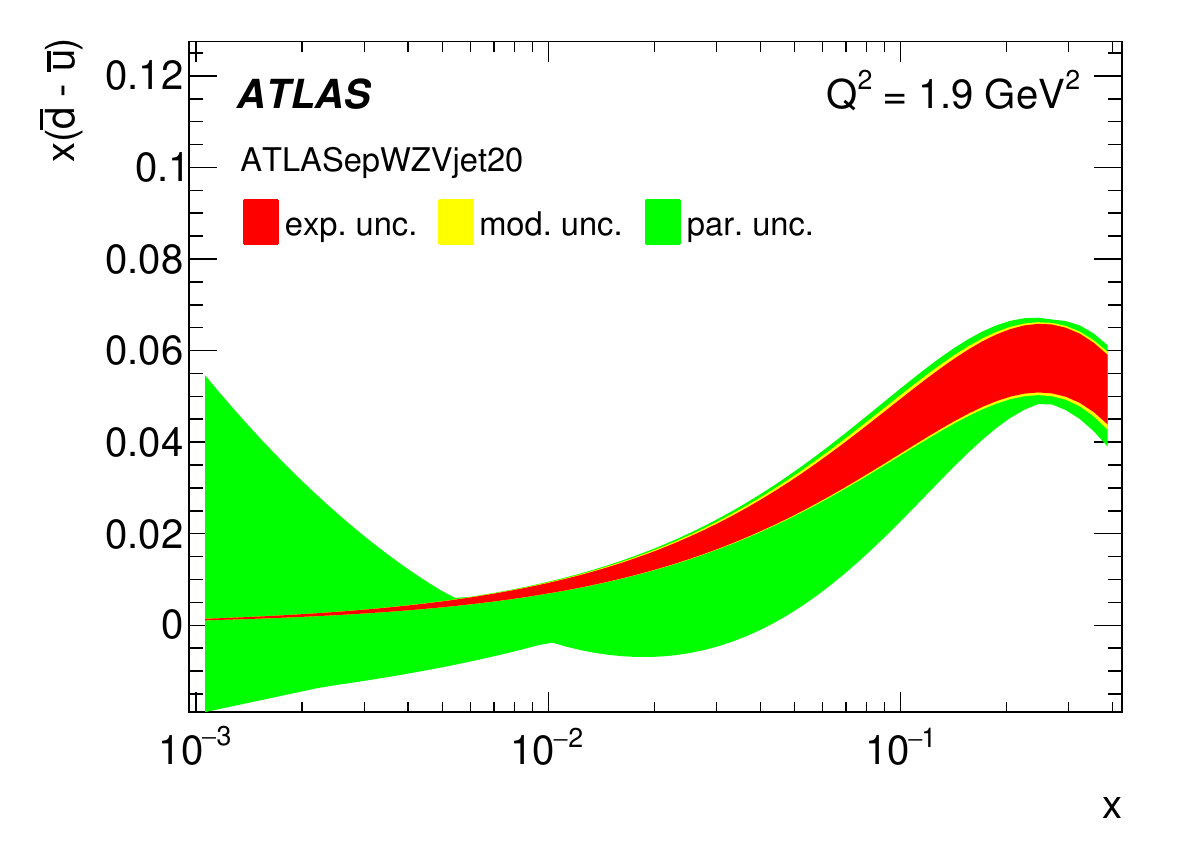}}
 
\caption{The $x(\dbar-\ubar)$ distribution evaluated at $Q^2 = 1.9~\GeV{}^2$ as a function of Bjorken $x$ (a) extracted from the ATLASepWZ20 (green) and ATLASepWZVjet20 (blue) fits with experimental and total uncertainties plotted separately, and (b) extracted from the ATLASepWZVjet20 fit only with experimental, model and parameterisation uncertainties shown separately in red, yellow and green, respectively.
}
\label{fig:dbarubarWJets}
\end{figure}
% End of text imported from the .//sections/Results/dbarubar.tex input file
% The next lines are included from the .//sections/Results/Chi2Scans.tex input file
To understand the effect of the different data sets on the high-$x$ \xdbar distribution, a scan of $\chi^2$ is performed through the parameter controlling the behaviour in this region, $C_{\dbar}$.\footnote{The other main contributor to the difference between the fits, $C_{\sbar}$, could equally be considered and would provide a similar insight as these two parameters are highly correlated.}
A high $C_{\dbar}$ value of $\sim 10$ corresponds to a lower \xdbar{} distribution at high $x$, as exhibited by the ATLASepWZ20 fit.
Conversely, a low $C_{\dbar}$ value of $\sim 2$ corresponds to the higher \xdbar{} distribution at high $x$ as exhibited by the ATLASepWZVjet20 fit.
 
In \Cref{fig:FullScan}, this scan is shown for each of the presented PDF fits, where the $\chi^2$ is evaluated as a function of the scanned parameter, $C_{\dbar}$.
At each point, all other parameters (including nuisance parameters associated with experimental uncertainties) are re-fitted and the minimum $\chi^2$ of the scan, $\chi^2_{\mathrm{min}}$, is subtracted for comparison between fits.
 
The $\chi^2$ of the ATLASepWZ20 fit is smallest at a value of $C_{\dbar} = 10 \pm 1$, whereas the $\chi^2$ of the ATLASepWZVjet20 fit is smallest at a lower $C_{\dbar} = 1.6 \pm 0.3$, corresponding to a higher $\xdbar$ distribution at $x \gtrsim 0.1$ consistent with the PDFs presented in \cref{subsec:DataFits}.
Another shallow minimum is observed for the ATLASepWZ20 fit at $C_{\dbar} \sim 3$, corresponding to a solution similar to that of the ATLASepWZVjet20 fit; however, it exhibits a $\chi^2$ approximately two units larger than in the best fit.
The ATLASepWZVjet20 fit fails to converge for values of $C_{\dbar} \gtrsim 12$ and no second minimum is observed.
 
In \Cref{fig:PartScan}, these $\chi^2$ distributions are decomposed into contributions from the HERA and ATLAS data.
These contributions include the partial, correlated and log penalty $\chi^2$, which are discussed in \cref{sec:theory}.
In each fit, the ATLAS data favour a low $C_{\dbar}$, including in the ATLASepWZ20 fit, where the overall result is a higher $C_{\dbar}$.
Similarly, the HERA data favour the higher $C_{\dbar}$ value exhibited by the ATLASepWZ20 fit.
The $V$~+~jets data provide sufficient constraining power in addition to the inclusive $W$ and $Z$ data to dominate the result and tightly constrain the $C_{\dbar}$ parameter to a low value, while the ATLASepWZ20 fit lacks the necessary information.
 
\begin{figure}
\centering
\subfloat[]{\includegraphics[width=.49\linewidth]{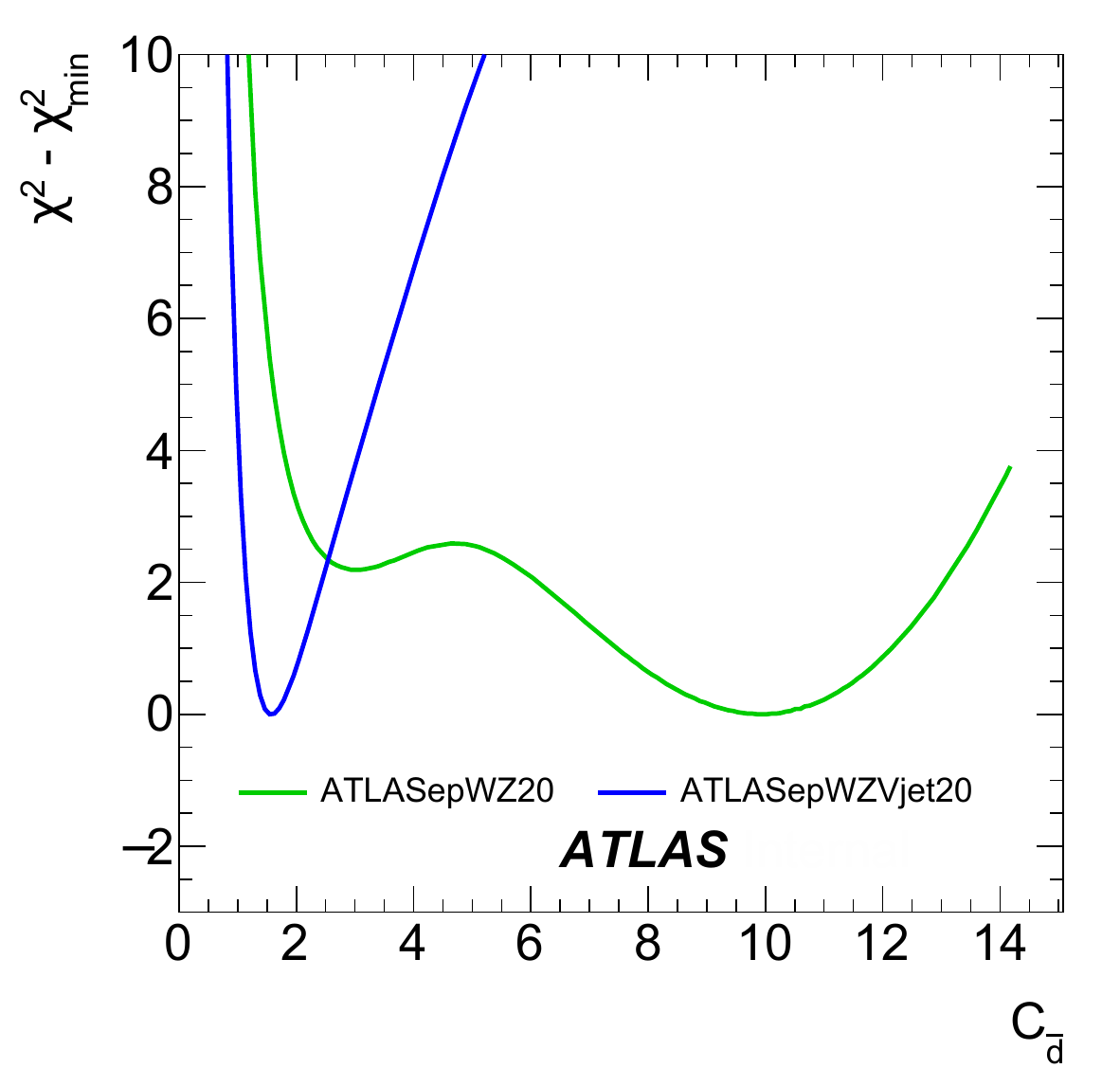}\label{fig:FullScan}}
\subfloat[]{\includegraphics[width=.49\linewidth]{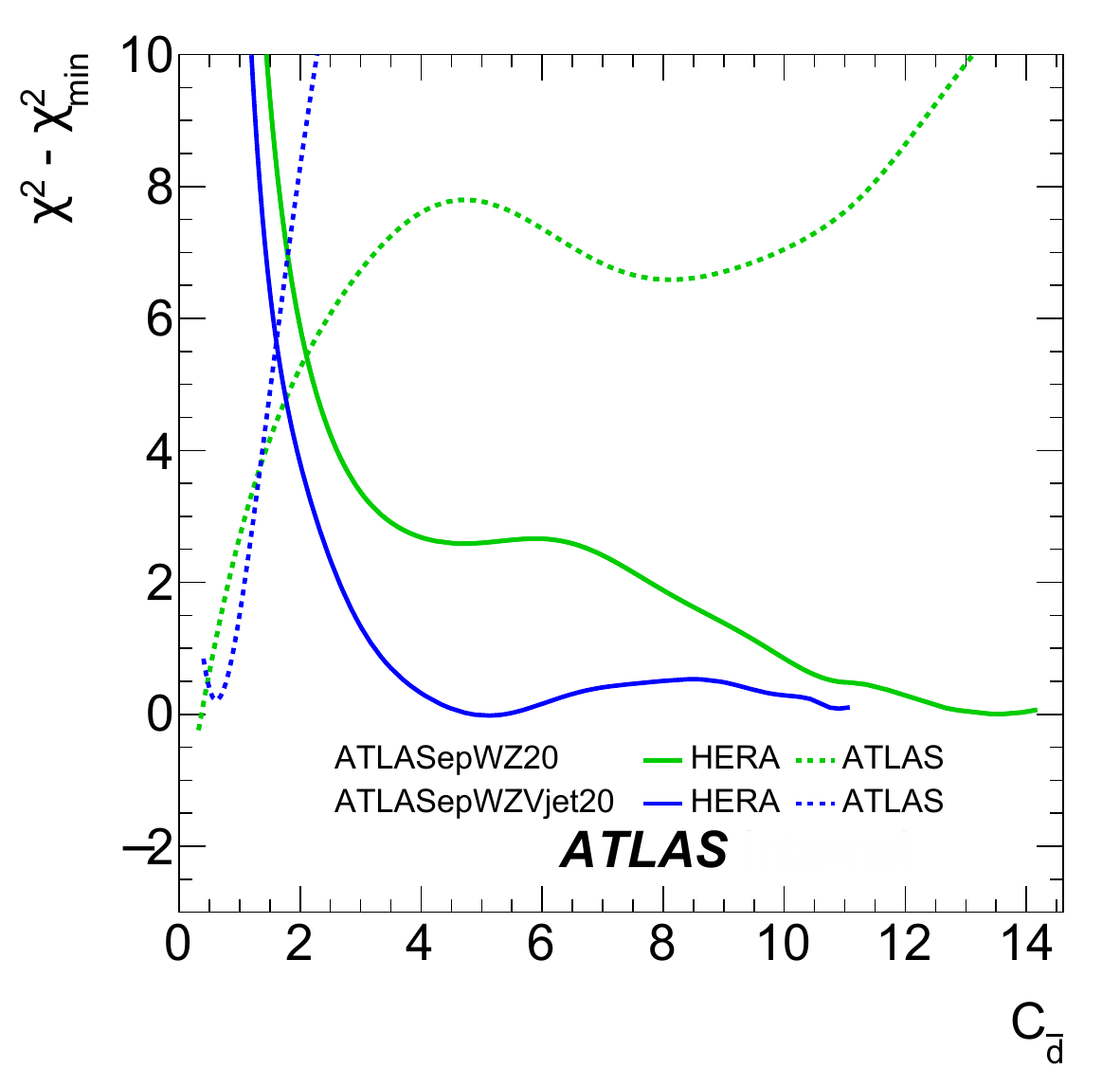}\label{fig:PartScan}}
\caption{The $\chi^2$ of the ATLASepWZ20 (green line) and ATLASepWZVjet20 (blue line) fits recorded as a function of the $C_{\dbar}$ fit parameter that determines the high-$x$ behaviour of the \xdbar{} PDF with $x\dbar \propto (1-x)^{C_{\dbar}}$. At each point, all other parameters are fitted along with the nuisance parameters corresponding to experimental systematic uncertainties, and the lowest recorded $\chi^2$ for each line shown, $\chi^2_{\mathrm{min}}$, subtracted.
Shown are (a) the total $\chi^2$ and (b) the $\chi^2$ separated into contributions from HERA (solid lines) and ATLAS (dashed lines) data.}
\end{figure}
% End of text imported from the .//sections/Results/Chi2Scans.tex input file
% The next lines are included from the .//sections/Results/Rs.tex input file
\subsection{Strange-quark density}
 
The fraction of the strange-quark density in the proton can be characterised by the quantity $R_s$, defined as the ratio
\begin{equation}\nonumber
R_s = \frac{s + \bar{s}}{\bar{u}+\bar{d}},
\end{equation}
which uses the sum of $\bar{u}$ and $\bar{d}$ as a reference point for the strange-sea density.
 
Before the first LHC precision $W$ and $Z$ boson data, it was widely
assumed, motivated by previous analyses of dimuon production in neutrino scattering \cite{PhysRevLett.99.192001,PhysRevD.91.094002,Kayis_Topaksu_2011,SAMOYLOV2013339,PhysRevD.64.112006}, that the strange sea-quark density is suppressed equally for all $x$
relative to the up and down sea over the full range of $x$.
Best fits to this neutrino scattering data resulted in a value of $R_s \sim 0.5$ at $Q^2=1.9~\GeV{}^2$ \cite{CT18,MMHT14,ABMP16,NNPDFStrange}.
 
The QCD analysis of the inclusive $W$ and $Z$ measurements by ATLAS which formed the ATLASepWZ16 PDF set
led to the observation that strangeness is unsuppressed at
low $x$ ($\lesssim 0.023$) for $Q^2 = 1.9~\GeV^2$.\footnote{
$x = 0.013$ is evaluated at $Q^2 = m_{Z}^2$ as this corresponds to the mean $x$ of inclusive $Z$ production at $\rts = 7~\TeV$, $\langle x_{Z} \rangle = m_{Z}/2E_{p}$. At the scale $Q^2 = 1.9~\GeV^2$, this corresponds to $x = 0.023$ through DGLAP evolution.}
This was the case for the ATLASepWZ16 fit for every parameterisation variation used.
Furthermore, a Hessian profiling exercise of the global PDFs MMHT14 \cite{MMHT14} and CT14 \cite{CT14} demonstrated that the data constrain and increase the ratio of the strange to the total up and down sea \cite{STDM-2012-20}.
Although profiling the PDFs does not necessarily give the same result as including the data in a fit, this effect is indeed found when the data is added to the CT18 fit, resulting in the CT18A set of PDFs \cite{CT18}.
It is therefore of particular interest to check the impact of the new $V$+ jets data on the strange-quark density.
 
The $R_s$ distribution plotted as a function of $x$ evaluated at $Q^2 = 1.9\text{~GeV}^2$ is shown in \Cref{fig:RsWJets}, with a comparison between ATLASepWZVjet20 and ATLASepWZ20 showing the direct effect of the $V$~+~jets data, and with the experimental, model and parameterisation uncertainties of ATLASepWZVjet20 shown separately.
The effect of the $V$~+~jets data is most significant in the kinematic region $x > 0.02$, where the uncertainty is significantly reduced.
Whereas the $R_s$ distribution of the ATLASepWZ20 PDFs maintained an unsuppressed strange-quark density over a wide range in $x$, the ATLASepWZVjet20 PDFs exhibit an $R_s$ distribution falling from near-unity at $x\sim 0.01$ to approximately $0.5$ at $x=0.1$, driven by the increase in the high-$x$ \dbar{} PDF and the complementary decrease in the high-$x$ \sbar{} PDF shown in \cref{subsec:DataFits}.
At low $x \lesssim 0.023$ and $Q^2 = 1.9~\GeV^2$, the fit with the $V$~+~jets data maintains an unsuppressed strange-quark density compatible with the ATLASepWZ16 fit.
Fitted values of $R_s$, evaluated at $x=0.023$ and $Q^2 = 1.9~\GeV{}^2$, are given in \cref{tab:Rs}.
 
{\renewcommand{\arraystretch}{1.5}
\begin{table}[h!]
\centering
\begin{tabular}{c c c c c}
\hline
\hline
& & \multicolumn{3}{c}{Uncertainties}\\
Fit & $R_{s}$ & Experimental & Model & Parameterisation \\
\hline
ATLASepWZ16 & $1.13$ & $0.05$ & $0.03$ & $^{+0.01}_{-0.06}$\\
ATLASepWZ20 & $1.13$ & $0.06$ & $0.03$ & $^{+0.09}_{-0.17}$\\
ATLASepWZVjet20 & $0.99$ & $0.04$ & $^{+0.05}_{-0.06}$ & $^{+0.14}_{-0.05}$\\
\hline\hline
\end{tabular}
\caption{Fitted values of $R_{s} = (s+\sbar)/(\ubar+\dbar)$, evaluated at $x = 0.023$ and $Q^{2} = 1.9~\GeV^{2}$, for each of the investigated fits compared with the ATLASepWZ16 result.}
\label{tab:Rs}
\end{table}
}
 
\begin{figure}
\centering
\subfloat[]{\includegraphics[width=.49\linewidth]{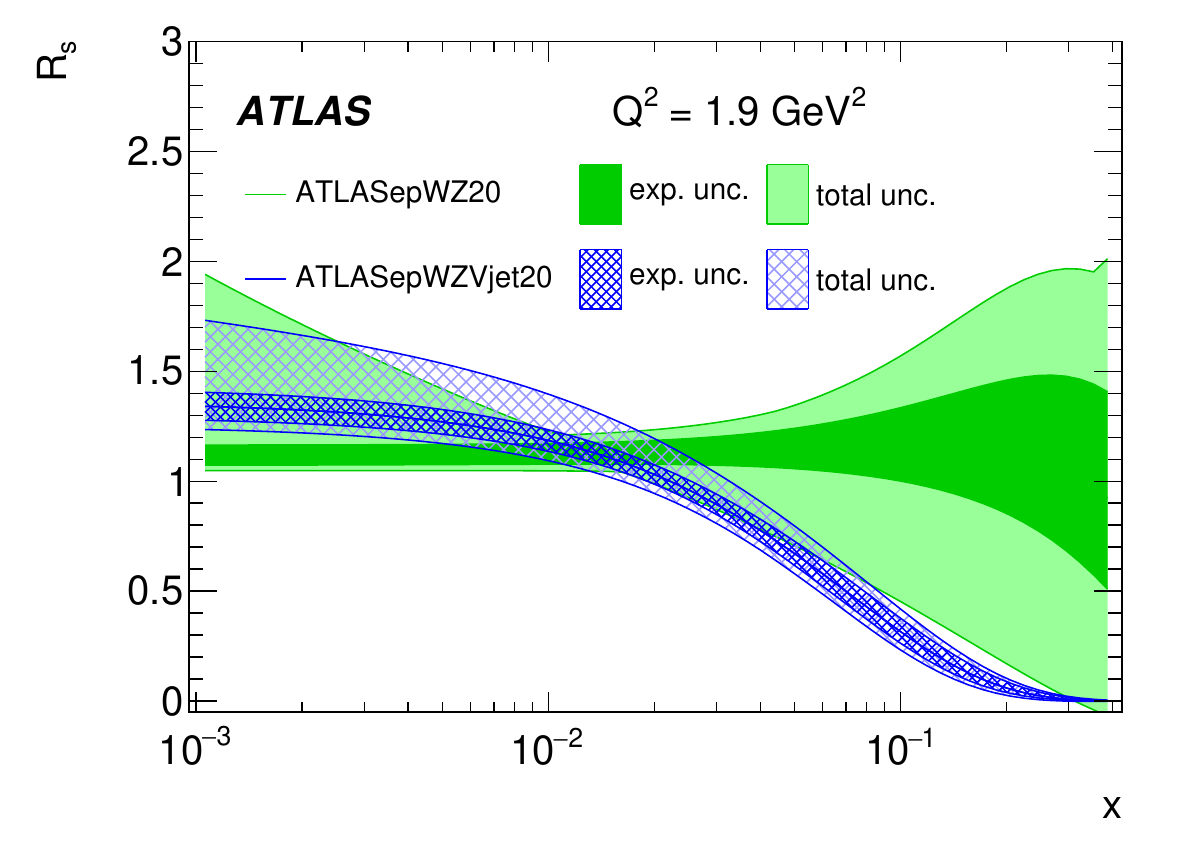}}
\subfloat[]{\includegraphics[width=.49\linewidth]{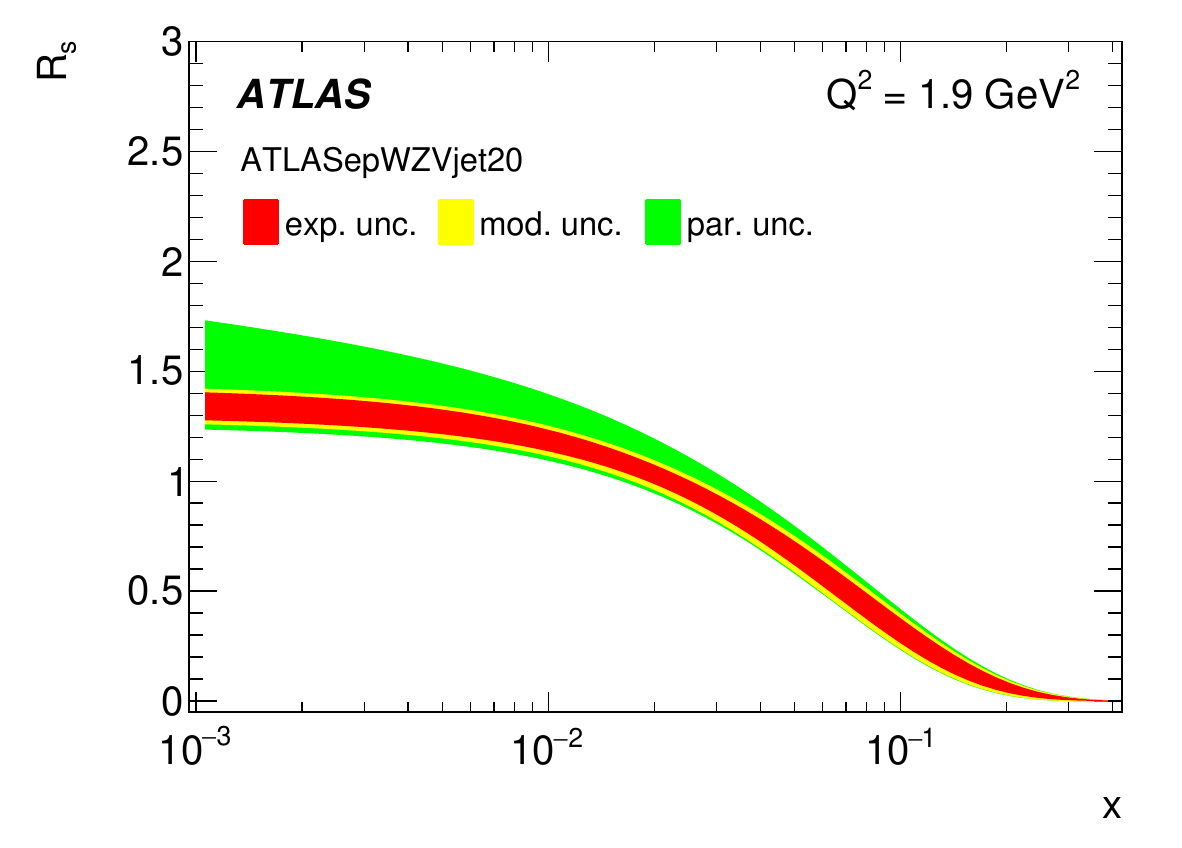}}
\caption{
The $R_{s} = (s + \sbar)/(\ubar + \dbar)$ distribution, evaluated at $Q^2 = 1.9~\GeV{}^2$, (a) extracted from the ATLASepWZ20 (green) and ATLASepWZVjet20 (blue) fits with experimental and total uncertainties plotted separately, and (b) extracted from the ATLASepWZVjet20 fit only with experimental, model and parameterisation uncertainties shown separately in red, yellow and green, respectively.
}
\label{fig:RsWJets}
\end{figure}
 
% End of text imported from the .//sections/Results/Rs.tex input file
% The next lines are included from the .//sections/Results/GlobalComparison.tex input file
\subsection{Comparison with global PDFs}
 
The ATLASepWZVjet20 $R_s$ distribution is shown in \Cref{fig:RsSummaryFull} in comparison with the global PDF sets ABMP16 \cite{ABMP16}, CT18, CT18A \cite{CT18}, MMHT14 \cite{MMHT14} and NNPDF3.1 \cite{NNPDF31}~\footnote{All these comparisons are done only at NNLO, and not at NLO.}.
An additional comparison in the figures is made with a recent update of the NNPDF3.1 fit with some additional data including the full ATLAS 7~\TeV{} data set labelled NNPDF3.1\_strange \cite{NNPDFStrange}.
Tension between the ATLASepWZVjet20 fit and the global analyses is reduced compared to the ATLASepWZ16 and ATLASepWZ20 PDF sets, but persists to multiple standard deviations in the range $10^{-2} \lesssim x \lesssim 10^{-1}$ for the global analyses which do not use the full ATLAS 7~\TeV{} data set. This is highlighted in summary plots of $R_s$ evaluated at $x=0.023$, $Q^2=1.9~\GeV{}^2$ and at $x=0.013$, $Q^2=m_{Z}^2$ in \Cref{fig:RsSummary}.
Better agreement is observed with the CT18A PDF set, which includes both the data used in the CT18 fit and the ATLAS 7~\TeV{} data, although tension remains with the NNPDF3.1\_strange PDF set, which also uses this data.
At high $x \gtrsim 0.02$, the $R_s$ distribution of the ATLASepWZVjet20 fit falls more steeply than the $R_s$ distribution in global analyses and is approximately zero at $x \gtrsim 0.2$.
The uncertainty of the NNPDF sets are large at $x > 0.3$ where the data give no constraint.
 
\begin{figure}
\centering
\subfloat[]{\includegraphics[width=.49\linewidth]{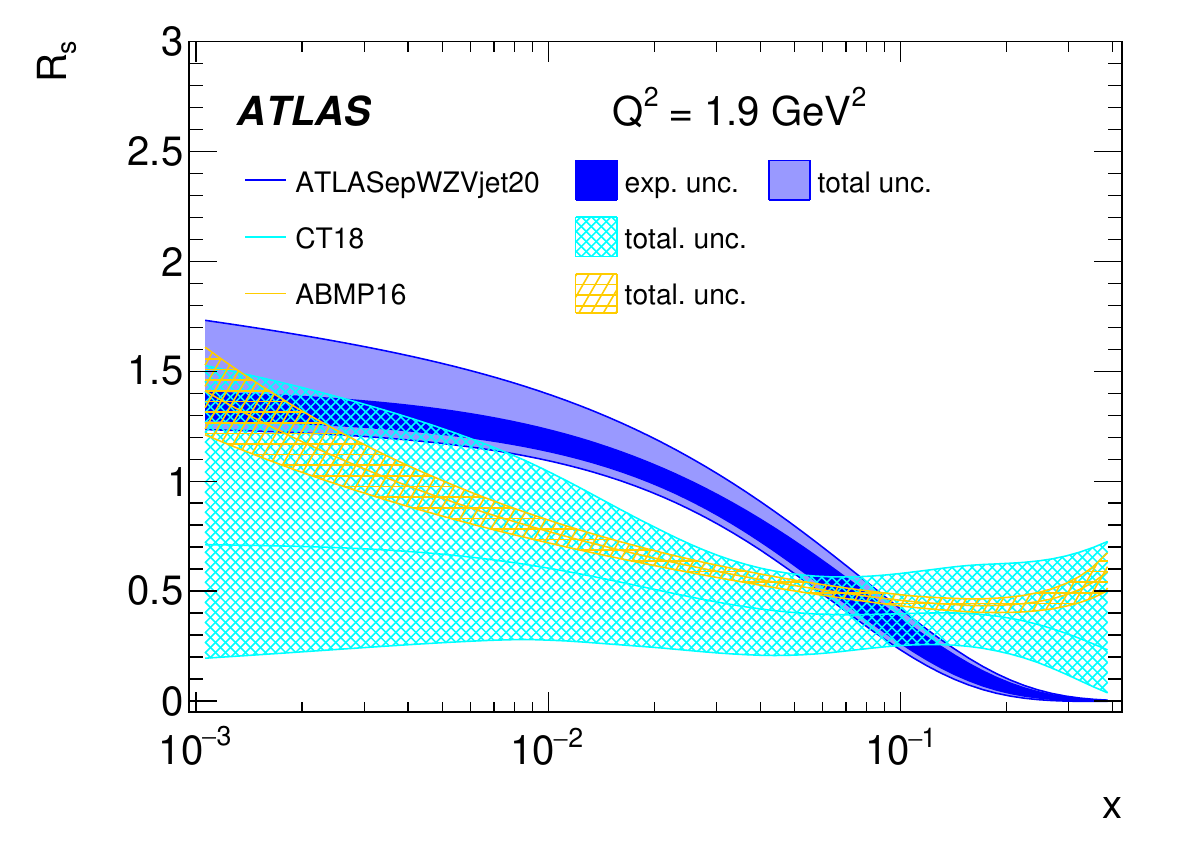}}
\subfloat[]{\includegraphics[width=.49\linewidth]{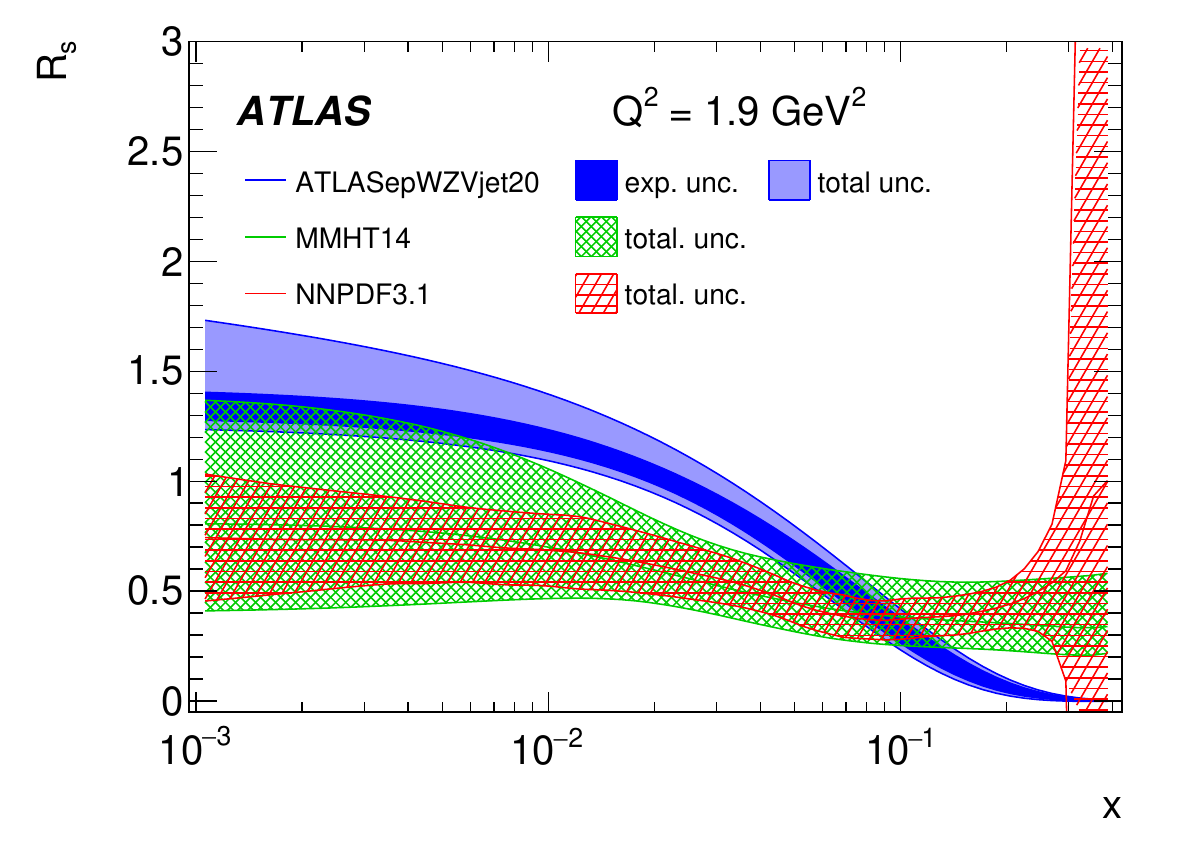}}
 
\subfloat[]{\includegraphics[width=.49\linewidth]{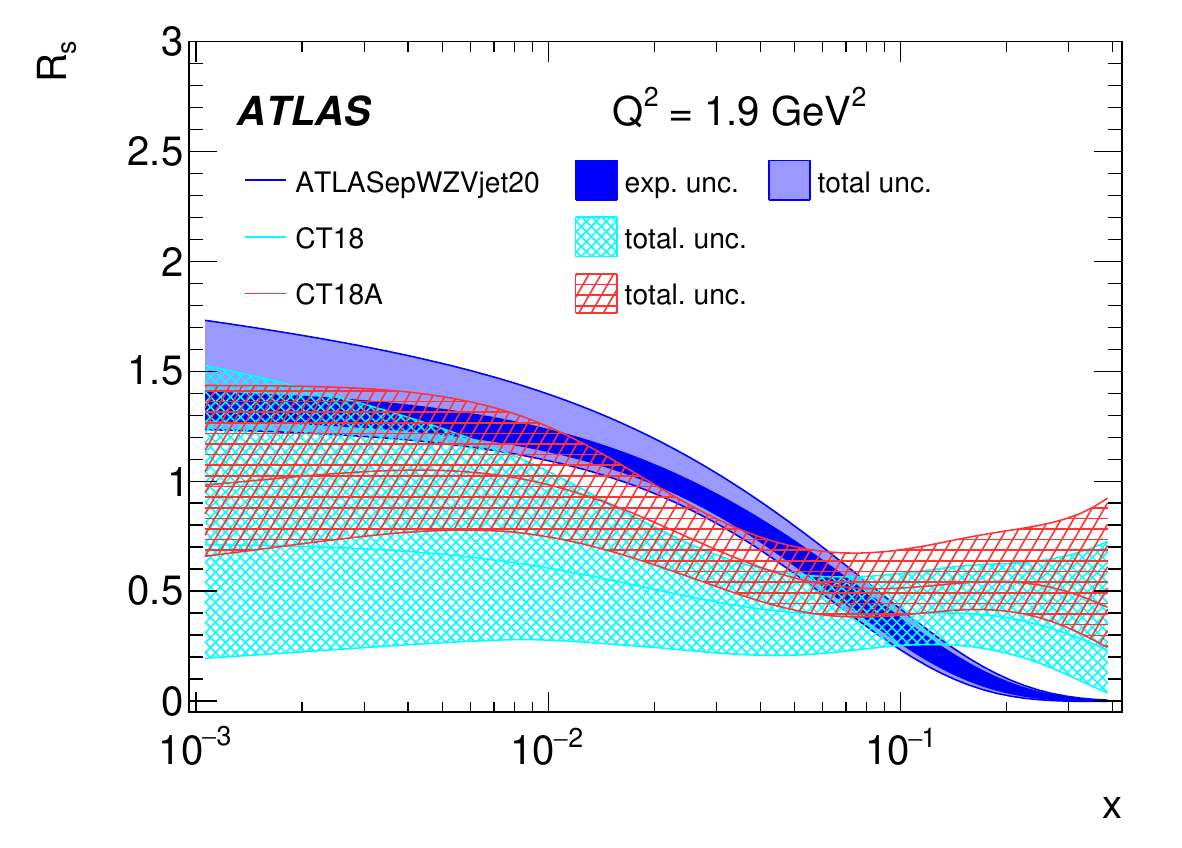}}
\subfloat[]{\includegraphics[width=.49\linewidth]{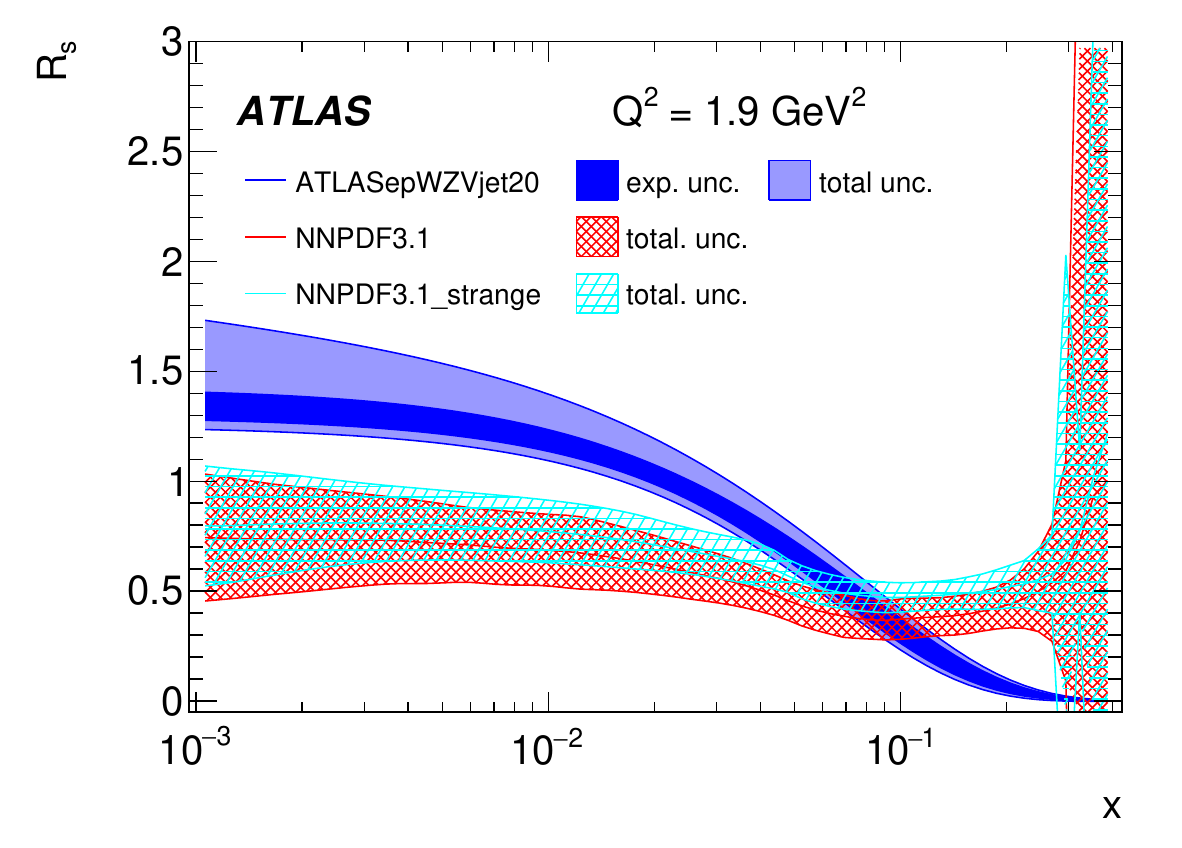}}
\caption{
The $R_s = (s + \sbar)/(\ubar+\dbar)$ distribution evaluated at $Q^2=1.9~\GeV{}^2$ as a function of Bjorken $x$, for the ATLASepWZVjet20 PDF set in comparison with global PDFs (a) ABMP16 and CT18, (b) MMHT14 and NNPDF3.1, and in additional comparisons with (c) CT18 and CT18A, and (d) NNPDF3.1 and NNPDF3.1\_strange \cite{CT18,MMHT14,NNPDF31,ABMP16,NNPDFStrange}. The experimental and total uncertainty bands are plotted separately for the ATLASepWZVjet20 results. Each global PDF set is taken at $\alphas(m_Z) = 0.1180$ except for ABMP16 which uses the fitted value $\alphas(m_Z) = 0.1147$. All global PDF uncertainty bands are at 68\% confidence level, evaluated for the CT18 PDFs through scaling by 1.645 as recommended by the PDF4LHC group \cite{Butterworth:2015oua}.
}
\label{fig:RsSummaryFull}
\end{figure}
 
\begin{figure}
\centering
\subfloat[]{\includegraphics[width=.49\linewidth]{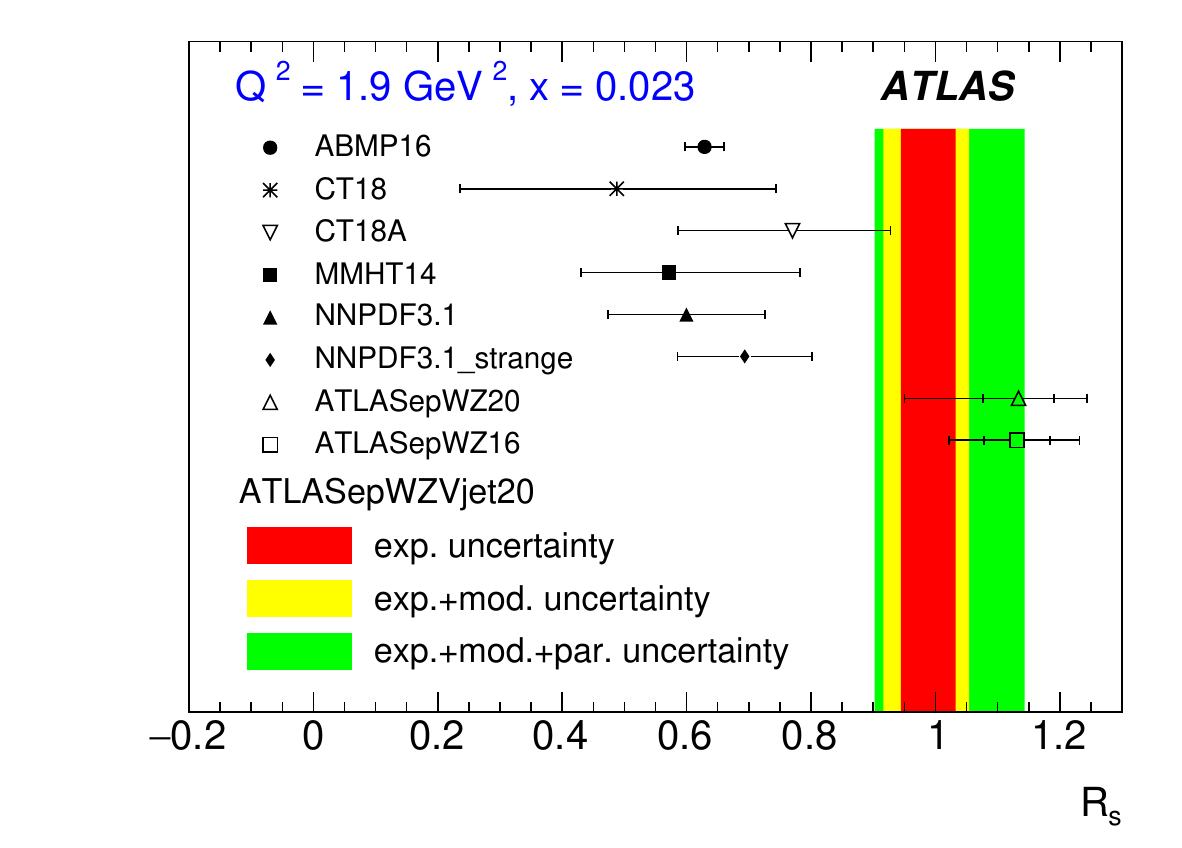}}
\subfloat[]{\includegraphics[width=.49\linewidth]{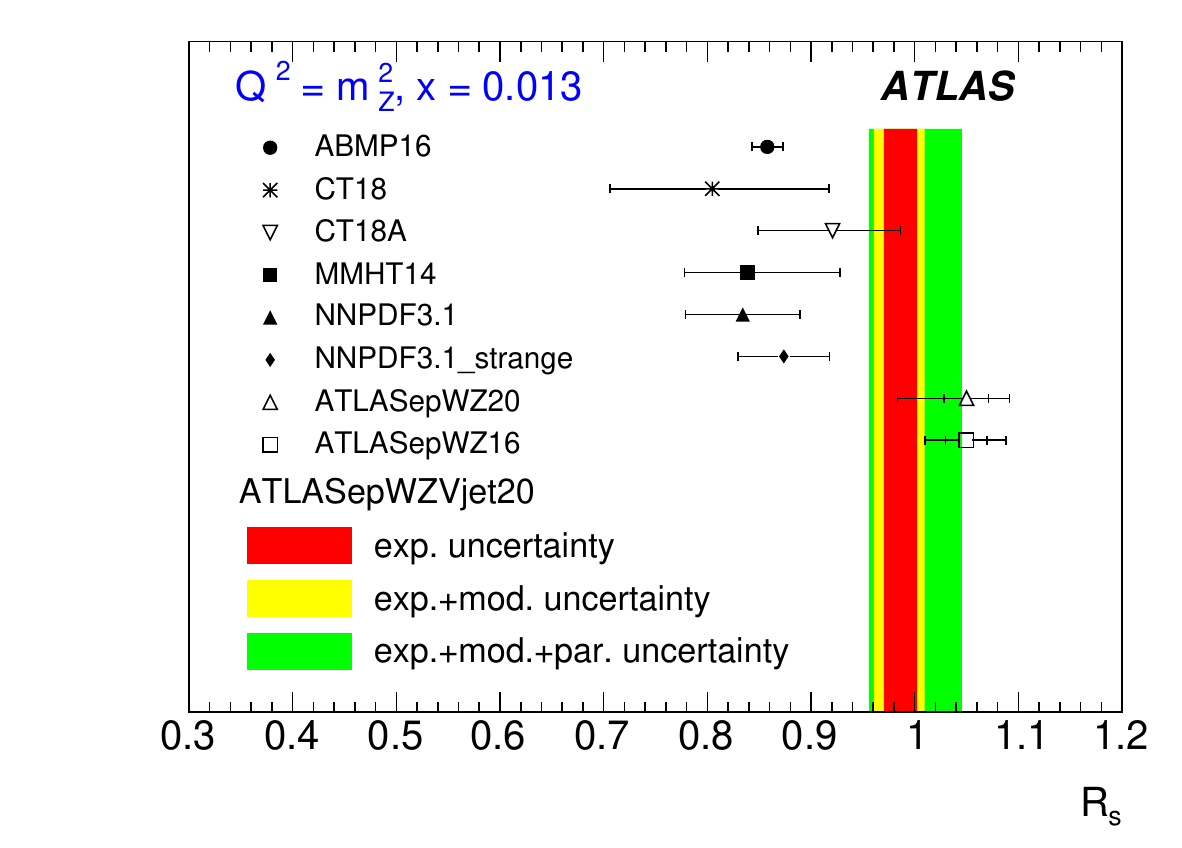}}
\caption{Summary plots of $R_s$ evaluated at (a) $x=0.023$ and $Q^2=1.9~\GeV{}^2$, and (b) $x=0.013$ and $Q^2=m_{Z}^{2}$, for the ATLASepWZVjet20 PDF set in comparison with global PDFs \cite{CT18,MMHT14,NNPDF31,ABMP16,NNPDFStrange}, and the ATLASepWZ16 and ATLASepWZ20 sets. The experimental, model and parameterisation uncertainty bands are plotted separately for the ATLASepWZVjet20 results. Each global PDF set is taken at $\alphas(m_Z) = 0.1180$ except for ABMP16 which uses the fitted value $\alphas(m_Z) = 0.1147$. All uncertainty bands are at 68\% confidence level, evaluated for the CT18 PDFs through scaling by 1.645 as recommended by the PDF4LHC group \cite{Butterworth:2015oua}.
}
\label{fig:RsSummary}
\end{figure}
 
In \Cref{fig:dbarubarSummary} the extracted $x(\dbar-\ubar)$ distribution
at $Q^2 = 1.9~\GeV{}^2$ is shown in comparison with the results
of the latest global PDF sets, all of which use E866 data.
The ATLASepWZVjet20 PDF set is consistent with these global PDF sets up to $x \sim 0.1$, but deviates from them for $x > 0.1$, where the $W$~+~jets and $Z$~+~jets data are most sensitive and demonstrate a preference for a higher $x\dbar$ distribution as discussed in \cref{subsec:DataFits}. A new result from SeaQuest/E906 Collaboration has recently become available \cite{Dove:2021ejl}, which may also be in tension with the E866 data.
Whereas the $R_s$ distribution of the CT18A and NNPDF3.1\_strange fits is affected by the ATLAS data and the tension between these and the ATLAS fits is reduced, as shown in \Cref{fig:RsSummaryFull,fig:RsSummary}, this is not replicated in the $x(\dbar-\ubar)$ distribution in either case.\\
 
\begin{figure}
\centering
\subfloat[]{\includegraphics[width=.49\linewidth]{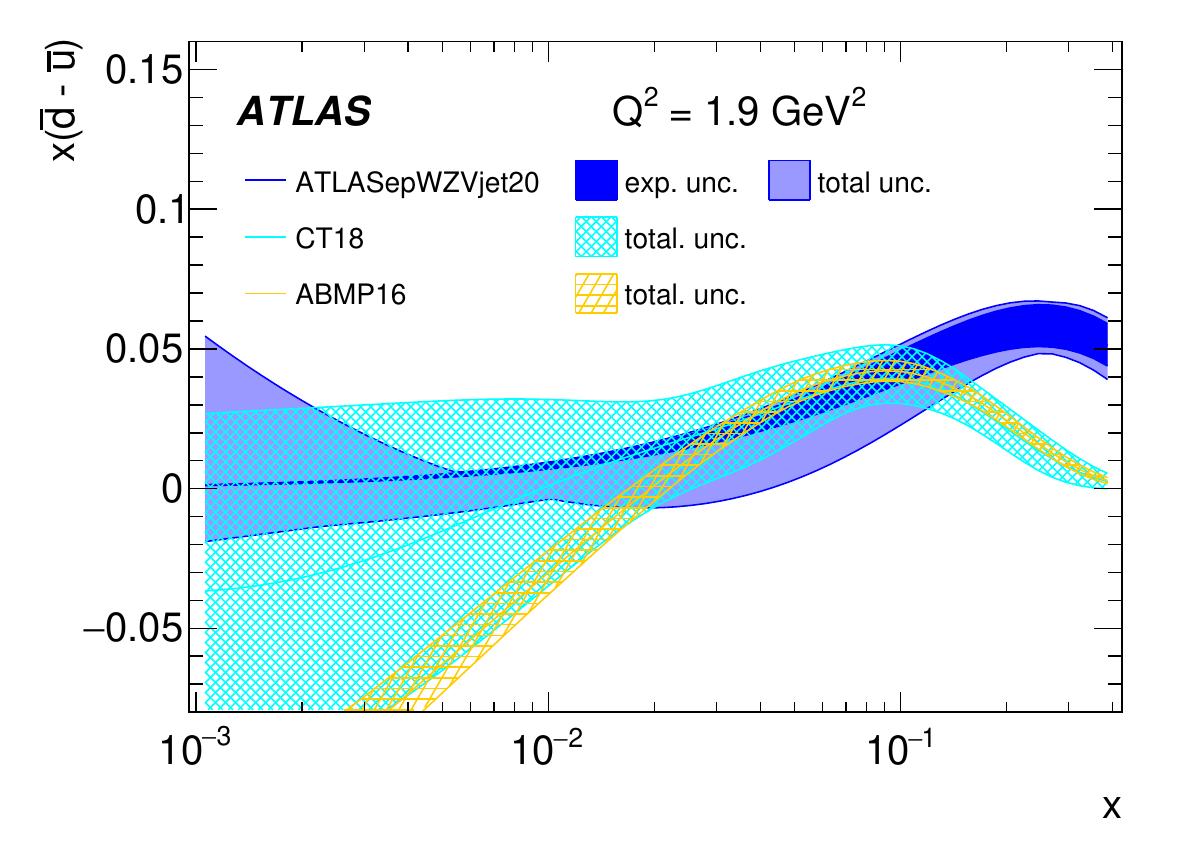}}
\subfloat[]{\includegraphics[width=.49\linewidth]{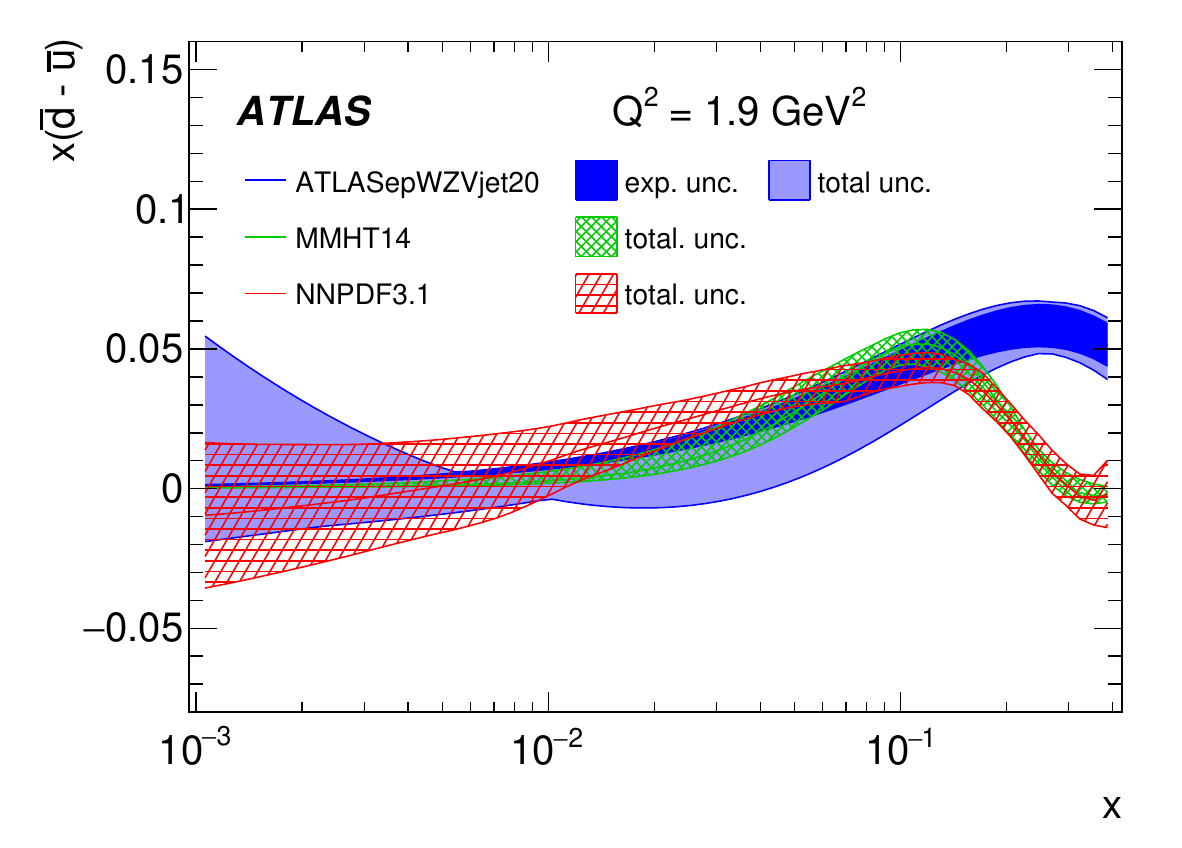}}
 
\subfloat[]{\includegraphics[width=.49\linewidth]{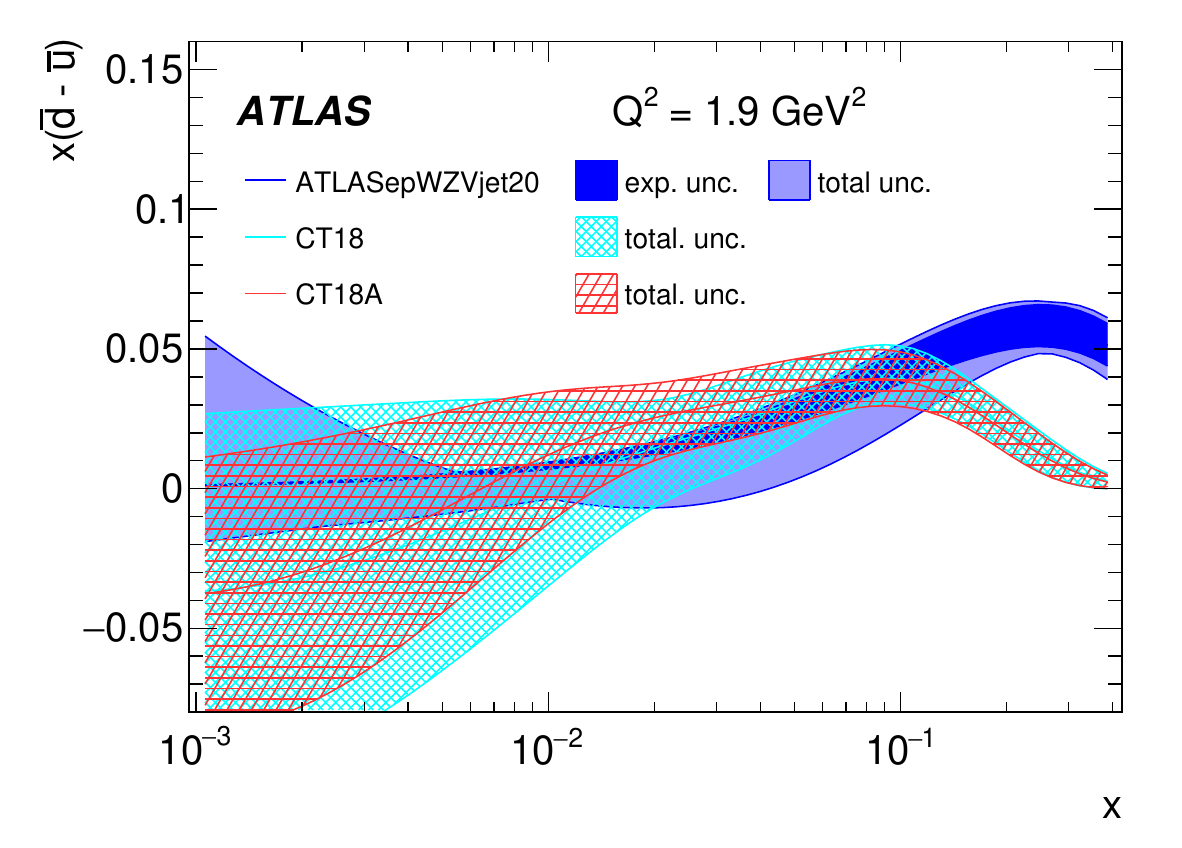}}
\subfloat[]{\includegraphics[width=.49\linewidth]{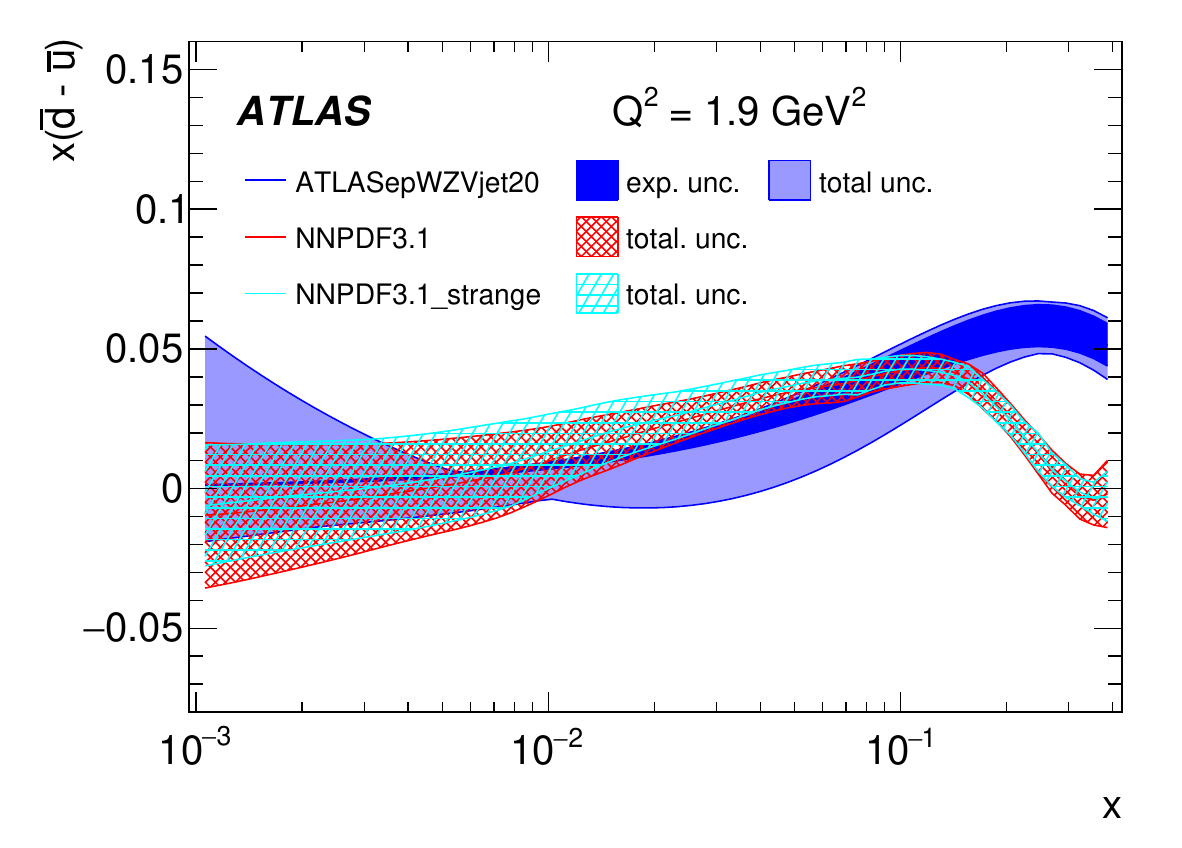}}
\caption{
The $x(\dbar-\ubar)$ distribution evaluated at $Q^2=1.9~\GeV{}^2$ as a function of Bjorken $x$, for the ATLASepWZVjet20 PDF set in comparison with global PDFs (a) ABMP16 and CT18, (b) MMHT14 and NNPDF3.1, and in additional comparisons with (c) CT18 and CT18A, and (d) NNPDF3.1 and NNPDF3.1\_strange \cite{CT18,MMHT14,NNPDF31,ABMP16,NNPDFStrange}. The experimental and total uncertainty bands are plotted separately for the ATLASepWZVjet20 result. Each global PDF set is taken at $\alphas(m_Z) = 0.1180$ except for ABMP16 which uses the fitted value $\alphas(m_Z) = 0.1147$. All global PDF uncertainty bands are at 68\% confidence level, evaluated for the CT18 PDFs through scaling by 1.645 as recommended by the PDF4LHC group \cite{Butterworth:2015oua}.
}
\label{fig:dbarubarSummary}
\end{figure}
% End of text imported from the .//sections/Results/GlobalComparison.tex input file
% End of text imported from the .//sections/Results.tex input file
 
\FloatBarrier
\section{Conclusion}
 
% The next lines are included from the .//sections/Conclusion.tex input file
This paper presents the impact of measurements, performed by the ATLAS experiment at the LHC, of vector-boson production in association with at least one jet on the parton distribution functions of the proton, resulting in a new ATLASepWZVjet20 PDF set.
The $V$ + jets data was obtained from $pp$ collisions at $\rts = 8~\TeV$ corresponding to approximately 20~\ifb\ of integrated luminosity.
The data were fitted along with the data sets used for the previous ATLASepWZ16 fit, i.e.\ the full combined
inclusive data set from HERA and the ATLAS inclusive $W$ and $Z$
production data recorded at $\rts = 7~\TeV$. For the new ATLASepWZVjet20 PDF set, correlations between all significant systematic
uncertainties across different data sets were considered.
 
The resulting PDF set is similar to the ATLASepWZ16 set for the up-type quarks and gluon.
The down and strange sea-quark distributions exhibit significantly smaller experimental and parameterisation uncertainties at high Bjorken $x$. As a result, the ratio of the strange-quark to light-quark densities, $R_s$, is better constrained and found to fall more steeply at high $x$. The $x(\bar{d} - \bar{u})$ difference is positive, in better agreement with the global PDF analyses which use E866 Drell--Yan data up to $x \sim 0.1$ but differs at higher values of $x$ by up to two standard deviations.
At low $x \lesssim 0.023$, the fit confirms the unsuppressed strange PDF as observed in the ATLASepWZ16 PDF set, while it maintains a positive $x(\dbar-\ubar)$ distribution at high $x$.
% End of text imported from the .//sections/Conclusion.tex input file

\section*{Acknowledgements}
 
% The next lines are included from the .//acknowledgements/Acknowledgements.tex input file

We thank CERN for the very successful operation of the LHC, as well as the
support staff from our institutions without whom ATLAS could not be
operated efficiently.
 
We acknowledge the support of
ANPCyT, Argentina;
YerPhI, Armenia;
ARC, Australia;
BMWFW and FWF, Austria;
ANAS, Azerbaijan;
SSTC, Belarus;
CNPq and FAPESP, Brazil;
NSERC, NRC and CFI, Canada;
CERN;
ANID, Chile;
CAS, MOST and NSFC, China;
Minciencias, Colombia;
MEYS CR, Czech Republic;
DNRF and DNSRC, Denmark;
IN2P3-CNRS and CEA-DRF/IRFU, France;
SRNSFG, Georgia;
BMBF, HGF and MPG, Germany;
GSRI, Greece;
RGC and Hong Kong SAR, China;
ISF and Benoziyo Center, Israel;
INFN, Italy;
MEXT and JSPS, Japan;
CNRST, Morocco;
NWO, Netherlands;
RCN, Norway;
MEiN, Poland;
FCT, Portugal;
MNE/IFA, Romania;
JINR;
MES of Russia and NRC KI, Russian Federation;
MESTD, Serbia;
MSSR, Slovakia;
ARRS and MIZ\v{S}, Slovenia;
DSI/NRF, South Africa;
MICINN, Spain;
SRC and Wallenberg Foundation, Sweden;
SERI, SNSF and Cantons of Bern and Geneva, Switzerland;
MOST, Taiwan;
TAEK, Turkey;
STFC, United Kingdom;
DOE and NSF, United States of America.
In addition, individual groups and members have received support from
BCKDF, CANARIE, Compute Canada and CRC, Canada;
COST, ERC, ERDF, Horizon 2020 and Marie Sk{\l}odowska-Curie Actions, European Union;
Investissements d'Avenir Labex, Investissements d'Avenir Idex and ANR, France;
DFG and AvH Foundation, Germany;
Herakleitos, Thales and Aristeia programmes co-financed by EU-ESF and the Greek NSRF, Greece;
BSF-NSF and GIF, Israel;
Norwegian Financial Mechanism 2014-2021, Norway;
NCN and NAWA, Poland;
La Caixa Banking Foundation, CERCA Programme Generalitat de Catalunya and PROMETEO and GenT Programmes Generalitat Valenciana, Spain;
G\"{o}ran Gustafssons Stiftelse, Sweden;
The Royal Society and Leverhulme Trust, United Kingdom.
 
The crucial computing support from all WLCG partners is acknowledged gratefully, in particular from CERN, the ATLAS Tier-1 facilities at TRIUMF (Canada), NDGF (Denmark, Norway, Sweden), CC-IN2P3 (France), KIT/GridKA (Germany), INFN-CNAF (Italy), NL-T1 (Netherlands), PIC (Spain), ASGC (Taiwan), RAL (UK) and BNL (USA), the Tier-2 facilities worldwide and large non-WLCG resource providers. Major contributors of computing resources are listed in Ref.~\cite{ATL-SOFT-PUB-2021-003}.
 
% End of text imported from the .//acknowledgements/Acknowledgements.tex input file

\clearpage
\appendix
\part*{Appendix}
\addcontentsline{toc}{part}{Appendix}

% The next lines are included from the .//sections/Correlations.tex input file
\section{Correlations between data sets}\label{app:correlations}
 
% The next lines are included from the .//sections/VJetsCorrelationsDetails.tex input file
The correlation model used for the ATLAS data is summarised in \cref{tab:systs}, where the labels used are the same as those in the HEPData entries of the respective ATLAS $W$~+~jets \cite{STDM-2016-14,1635273} and $Z$~+~jets \cite{STDM-2016-11,1744201} publications.
 
% The next lines are included from the .//figures/SystematicsTable.tex input file
\begin{table}[h!]
\centering
\resizebox{\textwidth}{!}{
\begin{tabular}{l l l l}
\hline
\hline
Systematic Uncertainty & 7~\TeV{} inclusive $W, Z$ & 8~\TeV{} $W$ + jets & 8~\TeV{} $Z$ + jets\\
\hline
\multirow{12}{*}{Jet energy scale \cite{PERF-2012-01}}      & \multirow{10}{*}{*}        & JetScaleEff1     & ATL\_JESP1 \\ \cline{3-4}
&                            & JetScaleEff2     & ATL\_JESP2 \\ \cline{3-4}
&                            & JetScaleEff3     & ATL\_JESP3 \\ \cline{3-4}
&                            & JetScaleEff4     & ATL\_JESP4 \\ \cline{3-4}
&                            & JetScaleEff5     & ATL\_JESP5 \\ \cline{3-4}
&                            & JetScaleEff6     & ATL\_JESP6 \\ \cline{3-4}
&                            & JetScaleEta1     & ATL\_JESP7 \\ \cline{3-4}
&                            & JetScaleEta2     & ATL\_JESP8 \\ \cline{3-4}
&                            & JetScaleHighPt   & ATL\_JESP9 \\ \cline{3-4}
&                            & JetScaleMC       & ATL\_JESP10\\ \cline{2-4}
& JetScaleNPV                & JetScalePileup1  & ATL\_PU\_OffsetNPV\\ \cline{2-4}
& JetScaleMu                 & JetScalePileup2  & ATL\_PU\_OffsetMu\\ \cline{1-4}
Jet punchthrough \cite{PERF-2012-01}                 & -                          & JetScalepunchT   & ATL\_PunchThrough\\ \cline{1-4}
Jet resolution \cite{PERF-2012-01}                   & JetRes                     & JetResolution10  & ATL\_JER\\ \cline{1-4}
Jet flavour composition \cite{PERF-2012-01}          & -                          & JetScaleFlav1Known & ATL\_Flavor\_Comp\\ \cline{1-4}
Jet flavour response \cite{PERF-2012-01}             & -                          & JetScaleFlav2    & ATL\_Flavor\_Response \\ \cline{1-4}
Pile-up jet rejection (JVF) \cite{PERF-2014-03}                & -                          & JetJVFcut        & ATL\_JVF \\ \cline{1-4}
\MET scale \cite{PERF-2014-04}                      & MetScaleWen                & METScale         & - \\ \cline{1-4}
\multirow{2}{*}{\MET resolution \cite{PERF-2014-04}} & \multirow{2}{*}{MetRes}    & METResLong       & - \\ \cline{3-4}
&                            & METResTrans      & - \\ \cline{1-4}
Electron energy scale \cite{PERF-2013-05}                  & *                          & ElScaleZee       & ATL\_ElecEnZee \\ \cline{1-4}
Electron trigger efficiency \cite{ATLAS-CONF-2012-048}    & *                          & ElSFTrigger      & ATL\_Trig \\ \cline{1-4}
Electron reconstruction efficiency \cite{PERF-2013-03,PERF-2016-01} & *                      & ElSFReco         & ATL\_RecEff \\ \cline{1-4}
Electron identification efficiency \cite{PERF-2013-03,PERF-2016-01}         & *                          & ElSFId           & ATL\_IDEff \\ \cline{1-4}
Luminosity \cite{DAPR-2011-01,DAPR-2013-01}                       & *                          & LumiUncert       & ATL\_lumi\_2012\_8TeV \\ \cline{1-4}
$WW$ background cross section \cite{ATL-PHYS-PUB-2017-005}    & *                          & XsecDibos        & ATL\_WW\_xs \\ \cline{1-4}
Top background cross section \cite{ATL-PHYS-PUB-2017-007}      & *                          & XsecTop          & ATL\_ttbar\_xs \\ \cline{1-4}
\hline
\hline
\end{tabular}}
\caption{Correlation model for the systematic uncertainties of the ATLAS measurements of $W$~+~jets and $Z$~+~jets at 8~\TeV{} and inclusive $W$ and $Z$ at 7~\TeV{}.
Each row corresponds to one source of systematic uncertainty treated as fully correlated both within and across data sets.
The respective ATLAS publication describing each of these sources in detail is given.
Sources in different rows are uncorrelated with each other.
Each source is reported with the label of the systematic uncertainty used in the respective data set.
Where entries are omitted, that systematic uncertainty either does not exist for that data set (denoted by a `-') or it was left decorrelated from the others (denoted by a `*').
For the row of source ``\met resolution'', where the ``MetRes'' label in one column corresponds to ``MetResLong'' and ``MetResTrans'' labels in the other, this indicates that the ``MetResLong'' and ``MetResTrans'' uncertainties were combined in quadrature for one-to-one correlation.
Additional uncertainties not reported are treated as uncorrelated between different data sets.}
\label{tab:systs}
\end{table}
% End of text imported from the .//figures/SystematicsTable.tex input file
 
Systematic uncertainties related to the jet energy scale, jet energy resolution, rejection of jets from pile-up (JVF), missing transverse momentum (\MET) scale, \MET resolution, electron energy scale and electron energy resolution are taken to have a one-to-one correlation between the 8~\TeV{} data sets.
Additionally, systematic uncertainties related to electron efficiency scale factors (trigger, reconstruction and isolation) as well as luminosity are considered fully correlated between data sets of the same centre-of-mass energy, but are uncorrelated between 7~\TeV{} and 8~\TeV{} data.
The systematic uncertainty related to the top-background cross section in $W$~+~jets data is taken as correlated with the top-quark pair production cross-section uncertainty in the $Z$~+~jets data as this is the largest of the top-quark-related background contributions in both data sets.
Similarly, the systematic uncertainty related to the total diboson cross section in the $W$~+~jets data is taken as correlated with the systematic uncertainty in the $Z$~+~jets data related to the production of two $W$ bosons, as this is the highest background contribution.
In contrast, systematic uncertainties related to the multijet background are estimated independently in each measurement, and are therefore left uncorrelated.\\
The two systematic uncertainties in each of the $W$~+~jets and $Z$~+~jets spectra related to unfolding (labelled in HEPData as ``UnfoldOtherGen'' and ``UnfoldReweight'' in the $W$~+~jets data, and ``ATL\_unfold\_Data'' and ``ATL\_unfold\_MC'' in the $Z$~+~jets data)\footnote{``Unfolding'' here refers to the procedure used to correct reconstructed data to particle-level cross sections.} are fully decorrelated between spectra and bins within a single spectrum (in addition to the aforementioned statistical uncertainties) as they contain a large statistical component in both data sets owing to MC simulation statistics.
Treating this source of uncertainty as correlated between all $W$~+~jets bins, for example, increases the $\chi^2$ by approximately 200 for 30 data points, despite insignificant changes to the resulting PDFs.\\
Two systematic uncertainties in the $W$~+~jets data set related the \MET resolution are summed into a single component for one-to-one correlation with the \MET resolution systematic uncertainty in the 7~\TeV{} data set. A further ten systematic uncertainties in the $V$~+~jets data correspond to a single component related to the jet energy scale in the 7~\TeV{} data; it is preferred to keep the more detailed model of ten sources in the 8~\TeV{} data as they have a significant impact on the $V$~+~jets measurements and the impact of the single source in the 7~\TeV{} data is small.\\
Additionally, two systematic uncertainties related to pile-up dependence of the jet energy scale, one systematic uncertainty related to jet energy resolution and one further related to the \MET{} scale, are also correlated between 7~\TeV{} and 8~\TeV{} data sets, for a total of five correlated components.
Cross-checks have been performed demonstrating that alternative models, for example partially correlating the luminosity uncertainties between 7 and 8~\TeV{} data or leaving all systematic uncertainties uncorrelated between 7 and 8~\TeV{} data and using combined 7~\TeV{} $W$ and $Z$ data, provide similar resulting PDFs.
% End of text imported from the .//sections/VJetsCorrelationsDetails.tex input file
% End of text imported from the .//sections/Correlations.tex input file
 
\printbibliography
\clearpage % ATLAS Collaboration author list
% Reference date of STDM-2019-18 is 2020-10-07
% Author list last updated on date 09-FEB-22
% Data extracted on 09-Feb-2022 for paper reference STDM-2019-18
% at 5:24pm
 
\begin{flushleft}
\hypersetup{urlcolor=black}
{\Large The ATLAS Collaboration}

\bigskip

\AtlasOrcid[0000-0002-6665-4934]{G.~Aad}$^\textrm{\scriptsize 101}$,    
\AtlasOrcid[0000-0002-5888-2734]{B.~Abbott}$^\textrm{\scriptsize 127}$,    
\AtlasOrcid[0000-0002-7248-3203]{D.C.~Abbott}$^\textrm{\scriptsize 102}$,    
\AtlasOrcid[0000-0002-2788-3822]{A.~Abed~Abud}$^\textrm{\scriptsize 36}$,    
\AtlasOrcid[0000-0002-1002-1652]{K.~Abeling}$^\textrm{\scriptsize 53}$,    
\AtlasOrcid[0000-0002-2987-4006]{D.K.~Abhayasinghe}$^\textrm{\scriptsize 93}$,    
\AtlasOrcid[0000-0002-8496-9294]{S.H.~Abidi}$^\textrm{\scriptsize 29}$,    
\AtlasOrcid[0000-0002-8279-9324]{O.S.~AbouZeid}$^\textrm{\scriptsize 40}$,    
\AtlasOrcid{N.L.~Abraham}$^\textrm{\scriptsize 155}$,    
\AtlasOrcid[0000-0001-5329-6640]{H.~Abramowicz}$^\textrm{\scriptsize 160}$
% ...
\bigskip
\\

% ...
$^{29}$Physics Department, Brookhaven National Laboratory, Upton NY; United States of America.\\
% ...
$^{36}$CERN, Geneva; Switzerland.\\
% ...
$^{40}$Niels Bohr Institute, University of Copenhagen, Copenhagen; Denmark.\\
% ...
$^{53}$II. Physikalisches Institut, Georg-August-Universit\"{a}t G\"ottingen, G\"ottingen; Germany.\\
% ...
$^{93}$Department of Physics, Royal Holloway University of London, Egham; United Kingdom.\\
% ...
$^{101}$CPPM, Aix-Marseille Universit\'e, CNRS/IN2P3, Marseille; France.\\
$^{102}$Department of Physics, University of Massachusetts, Amherst MA; United States of America.\\
% ...
$^{127}$Homer L. Dodge Department of Physics and Astronomy, University of Oklahoma, Norman OK; United States of America.\\
% ...
$^{155}$Department of Physics and Astronomy, University of Sussex, Brighton; United Kingdom.\\
% ...
$^{160}$Raymond and Beverly Sackler School of Physics and Astronomy, Tel Aviv University, Tel Aviv; Israel.\\
% ...

\end{flushleft}

% Created with Glance <Atlas.Glance@cern.ch>

\end{document}